\documentclass[12pt]{article}
\pdfoutput=1
\usepackage{epsf,amsfonts,amssymb,epsfig,amsmath,graphics,cite}
\usepackage[dvipsnames]{xcolor}
\usepackage{graphicx,subfig}
\usepackage{booktabs}
\usepackage{cancel}
\usepackage{float}
\usepackage{verbatim}
\usepackage{mathrsfs}
\usepackage{empheq}
\usepackage{enumerate}
\usepackage{bbm}
\usepackage{soul}
\usepackage{slashed}

\usepackage[colorlinks=true
,urlcolor=blue
,anchorcolor=blue
,citecolor=blue
,filecolor=blue
,linkcolor=blue
,menucolor=blue
,linktocpage=true
]{hyperref}
 
\makeatletter

\newcommand{\Rmnum}[1]{\expandafter\@slowromancap\romannumeral #1@}
\makeatother

\newcommand{\nn}{\notag }
\def\be{\begin{equation}}
\def\ee{\end{equation}}

\newcommand{\ii}{\mathrm{i}}
\newcommand{\ex}{\mathrm{e}}
\newcommand{\diff}{\mathrm{d}}
\newcommand{\dd}{\mathrm{d}}
\newcommand{\R}{\mathbb{R}}
\newcommand{\Z}{\mathbb{Z}}
\newcommand{\vol}{\mathrm{vol}}
\newcommand{\Vol}{\mathrm{Vol}}
\newcommand{\C}{\mathbb{C}}

\newcommand{\abs}[1]{\left\lvert #1 \right\rvert}

\newcommand{\mc}[1]{\mathcal{#1}}

\newcommand{\hook}{\mathbin{\rule[.2ex]{.4em}{.03em}\rule[.2ex]{.03em}{.9ex}}}

\newcommand{\del}{\partial}

\renewcommand{\Re}{\operatorname{Re}}
\renewcommand{\Im}{\operatorname{Im}}
\newcommand{\cK}{\mc{K}}

\newcommand{\vpvar}{\rho}
\newcommand{\svar}{s}

\newcommand{\newc}{x}
\newcommand{\Zzeta}{Z}

\newcommand{\Pnew}{\mathfrak{p}}

\numberwithin{equation}{section}       

\begin{document}

\begin{titlepage}


\vskip 1cm

\begin{center}


{\Large \bf 
Localization and Attraction}
\vskip 1cm
{Pietro Benetti Genolini$^{\mathrm{a}}$, Jerome P. Gauntlett$^{\mathrm{b}}$,
Yusheng Jiao$^{\mathrm{b}}$,\\
\vskip 0.1cm
 Alice L\"uscher$^{\mathrm{c}}$  and James Sparks$^{\mathrm{c}}$}

\vskip 1cm

${}^{\mathrm{a}}$\textit{D\'epartment de Physique Th\'eorique, Universit\'e de Gen\`eve,\\
24 quai Ernest-Ansermet, 1211 Gen\`eve, Suisse\\}

\vskip 0.2cm

${}^{\mathrm{b}}$\textit{Blackett Laboratory, Imperial College, \\
Prince Consort Rd., London, SW7 2AZ, U.K.\\}

\vskip 0.2cm

${}^{\mathrm{c}}$\textit{Mathematical Institute, University of Oxford,\\
Andrew Wiles Building, Radcliffe Observatory Quarter,\\
Woodstock Road, Oxford, OX2 6GG, U.K.\\}

\vskip 0.2 cm

\end{center}

\vskip 0.5 cm

\begin{abstract}
\noindent  
We use equivariant localization to construct off-shell entropy functions for supersymmetric black holes in 
$\mathcal{N}=2$, $D=4$ gauged supergravity coupled to matter. This allows one to compute the black hole entropy without solving the supergravity equations of motion and provides a novel generalization of the attractor mechanism. We consider
magnetically charged black holes in $AdS_4$ which have an $AdS_2\times M_2$ 
near horizon geometry, where $M_2$ is a sphere or a spindle,
and we also obtain entropy functions for 
ungauged supergravity as a simple corollary. We derive analogous results for black strings and rings
in $D=5$ supergravity which have an $AdS_3\times M_2$ near horizon geometry, and in this setting
we derive an off-shell expression for the central charge of the dual $\mathcal{N}=(0,2)$, $d=2$~SCFT.
\end{abstract}

\end{titlepage}

\pagestyle{plain}
\setcounter{page}{1}
\newcounter{bean}
\baselineskip18pt

\tableofcontents


\section{Introduction}\label{sec:intro}

It has recently been shown
that supersymmetric solutions of supergravity with an R-symmetry Killing vector
are equipped with a set of equivariantly closed forms which can be constructed from bilinears in
the Killing spinor  \cite{BenettiGenolini:2023kxp,BenettiGenolini:2023yfe,BenettiGenolini:2023ndb}. 
The localization fixed point formula \cite{BV:1982,Atiyah:1984px} then allows one to compute
various physical observables without needing to solve the supergravity equations of motion. Furthermore, the formalism
naturally gives rise to results that are off-shell with, generically, a simple extremization required
in order to get the final result. Within the context of holography this latter feature 
is precisely dual to an extremization problem in the corresponding supersymmetric conformal field theory.

Here we further investigate the new equivariant calculus in the context of magnetically charged black holes of $\mathcal{N}=2$, $D=4$ gauged supergravity and black strings (or rings) of $D=5$ gauged supergravity.
More precisely, by analysing the near horizon geometries we are able to derive novel off-shell entropy functions, and furthermore provide a new perspective on the attractor mechanism.
In both cases we allow for 
arbitrary numbers of vector multiplets and we also consider hypermultiplets. 
Interestingly, off-shell entropy functions for the associated \emph{ungauged} supergravity theories will also follow as simple corollaries.  

We now elaborate a little, first discussing the case of black holes in $\mathcal{N}=2$, $D=4$ gauged supergravity.
By analysing spinor bilinears, we construct an off-shell entropy function for classes of supersymmetric $AdS_4$ black holes which have an $AdS_2\times M_2$ near horizon geometry and carry magnetic charge on $M_2$. We take $M_2$ to be topologically a two-sphere or a spindle with an azimuthal symmetry. The off-shell entropy function, explicitly given in~\eqref{enfnsspndle} below, depends on the magnetic charges, as well
as the values of certain scalar fields and a metric warp function at the two poles of the horizon $M_2$. 
Extremizing over this data then gives the entropy of the black hole solution, provided
the solution actually exists.\footnote{To be more precise: our extremization procedure gives a result for the $AdS_2\times M_2$ 
solution, assuming it exists. If, furthermore, this solution arises as the near horizon limit of an $AdS_4$ black hole solution, 
it will compute the entropy of the black hole.} This extremization problem is actually  
implied by extremizing the action itself, and hence our results 
can be viewed as a kind of supersymmetric refinement of the approach of Sen \cite{Sen:2005wa}. 

Our results also provide a new way of thinking about some aspects of the attractor mechanism \cite{Ferrara:1995ih,Strominger:1996kf,Ferrara:1996dd}, which 
can be viewed\footnote{In ungauged supergravity the attractor mechanism also reveals that the near horizon geometry is independent
of scalar moduli at the asymptotically flat boundary. By contrast in gauged supergravity the values of the scalars at the asymptotic $AdS$ boundary are generically fixed by the potential. The attractor mechanism has been extensively investigated in gauged supergravity;
see \cite{David:2023gee} for a recent discussion. For the black holes we focus on here, some early references include \cite{Cacciatori:2009iz,DallAgata:2010ejj,Hristov:2010ri}. A review of the attractor mechanism in ungauged $D=5$ supergravity is given in 
 \cite{Larsen:2006xm}.} as a way of characterizing the near horizon geometry as the end-point of a flow,
as well as providing a generalization thereof.
Consider, for example, static, magnetically charged black holes with spherical horizons and with
supersymmetry preserved via a topological twist. A standard procedure is to assume that the sphere\footnote{One can also consider constant curvature metrics on other Riemann surfaces, but as they do not have suitable Killing vectors our new equivariant techniques do not provide further insight.}
has a constant curvature metric which allows one to construct an ansatz for the metric and matter fields that only depends on a radial variable. Preservation of supersymmetry then leads to a set of BPS equations, which consist
of a set of first order flow
equations in the radial variable. Assuming one approaches an $AdS_2\times S^2$ near horizon solution, the scalars and warp factor are necessarily constant at the horizon and the flow equations lead to conditions
which express the warp factor and some of the constant scalars on the horizon in terms of the magnetic charges. Moreover, one
can, in effect, obtain the entropy of the putative black holes by carrying out an extremization of an entropy function
depending on the black hole horizon data, which has been shown, in the case of the STU model, to be dual to $\mathcal{I}$-extremization in the dual field theory \cite{Benini:2015eyy, Benini:2016rke}.

Here, instead, we consider a more general class of magnetically charged $AdS_4$ black holes.
For spherical horizons, we just assume that the near horizon geometry has an $AdS_2\times S^2$ factor with the $S^2$ having a Killing vector. That is, we do not assume the $S^2$ horizon has a constant curvature metric nor do we
assume that the $S^2$ at the $AdS_4$ boundary has such a metric. Furthermore, while the near horizon
geometry is static we do not need to assume the black hole is static.\footnote{This is particularly relevant 
in $D=5$ supergravity, discussed below, which can admit non-static black ring solutions, but with locally $AdS_3\times S^2$ horizons.} The presence of the Killing vector at the horizon is sufficient data to use our equivariant techniques to derive an off-shell entropy function, which is precisely the same as that arising in the standard attractor mechanism. 
Our approach therefore increases the scope of the attractor mechanism in that it covers all black hole solutions
that approach $AdS_2\times S^2$ in the near horizon, with the $S^2$ having a Killing vector. In particular, this allows for the possibility that the metric on the $S^2$ at the $AdS_4$ boundary is arbitrary, a possibility that was also investigated from
a different point of view in \cite{Bobev:2020jlb}.

We also consider black holes with spindle horizons \cite{Ferrero:2020laf}. Recall that a spindle is topologically a two-sphere but with
quantized conical deficit angles at the two poles. Such black holes are known to arise when acceleration is present, with
the conical deficits at the poles associated with the acceleration \cite{Ferrero:2020twa}. Since a spindle does not admit a constant curvature metric, the standard attractor mechanism cannot be deployed. However, assuming that the spindle has a Killing vector, we can again utilize equivariant localization to obtain an off-shell entropy function. 
Explicitly, for a spindle horizon, which locally looks like $\mathbb{R}^2/\mathbb{Z}_{n_\pm}$ at the two poles, 
the off-shell entropy function is given by 
\begin{equation}\label{enfnsspndle}
	S_{\rm BH} = \frac{\pi}{2G_4} \frac{1}{2b_0}\left[{\ii \mc{F}(\newc^I_+) - \sigma \ii \mc{F}(\newc^I_-})\right]\, .
\end{equation}
Here $\mc{F}$ is the prepotential of the $\mc{N}=2$, $D=4$ gauged supergravity, with $\ii\mc{F}$ \emph{real}, and $\sigma = \pm 1$ labels whether supersymmetry is preserved via a twist or an anti-twist \cite{Ferrero:2021etw}. 
The localization is performed with 
respect to an R-symmetry Killing vector bilinear $\xi$, which exists for any supersymmetric 
solution. Here we write
\begin{align}
\xi = b_0 \partial_{\varphi}\, ,
\end{align}
with $\partial_{\varphi}$ generating rotations of the horizon, 
 $\varphi$ having canonical period $2\pi$, and $b_0$ being a real parameter.  
The $\newc_\pm^I$ are real scalar quantities evaluated at the two poles of the spindle\footnote{The $\newc_\pm^I$ are also  values of equivariant first Chern classes at the two poles.}, where the index $I$ labels the number of vector multiplets present, and satisfy the constraints 
\begin{align}\label{introconstraints}
\xi_I \newc^I_+=2-2\frac{b_0}{n_+}\, , \quad \xi_I \newc^I_-=2+2\sigma\frac{b_0}{n_-}\, , \quad \Pnew^I=\frac{1}{2b_0}(\newc_+^I-\newc_-^I)\, .
\end{align}
Here the constants $\xi_I\in\mathbb{R}$ determine the Fayet--Iliopoulos (FI)  gauging of the theory, while $\Pnew^I$ are the magnetic charges. The black hole entropy 
is then obtained by extremizing \eqref{enfnsspndle} over the variables 
$\newc_+^I, \newc_-^I, b_0$, subject to the constraints \eqref{introconstraints}, where 
we hold the spindle data $n_\pm$, $\sigma$ and magnetic charges $\Pnew^I$ fixed.
Our results then provide a derivation of 
``gravitational block'' formulae of gauged supergravity that have been 
conjectured in the literature \cite{Hosseini:2019iad} (see also \cite{Hosseini:2021fge,Faedo:2021nub,Hosseini:2023ewi}). 

With regard to the results for $S^2$ horizons discussed above, we
 will find it most convenient to obtain them as a limiting case $b_0\rightarrow 0$ of the results for spindle horizons, by setting $n_\pm=1$ and
 $\sigma=1$, and we obtain
 \begin{equation}\label{enfnssphere}
S_{\rm BH} = \frac{\pi}{2G_4} \Pnew^I \frac{\partial \ }{\partial \newc^I} \ii\mc{F}(\newc^I) \, ,
\end{equation}
with the $\newc^I$ and magnetic charges constrained via $\xi_I \newc^I=2$ and $\xi_I \Pnew^I=-2$, in agreement 
with \cite{DallAgata:2010ejj} for purely electric gauging.    

Of particular interest is the $D=4$ STU model, an example of $\mc{N}=2$ gauged supergravity coupled to three vector multiplets, since it can be obtained as a consistent truncation of $D=11$ supergravity on $S^7$. 
For the STU model our results for the off-shell entropy functions associated with $AdS_2\times M_2$ horizons, where $M_2$ is a two-sphere or spindle, and three real scalar fields, precisely recover the known results in \cite{Benini:2015eyy}  and \cite{Faedo:2021nub, Ferrero:2021ovq, Couzens:2021cpk}, respectively. 
We can also take a $D=11$ perspective by uplifting on $S^7$ to obtain $AdS_2\times M_9$ solutions of $D=11$ supergravity,
with $M_9$ being an $S^7$ bundle over $M_2$ and having a GK geometry\cite{Kim:2006qu,Gauntlett:2007ts}. We then show that our new entropy functions \eqref{enfnsspndle} and \eqref{enfnssphere} are equivalent to the entropy functions and gravitational block formulae
that were derived within the context of GK geometry, using a higher-dimensional supergravity perspective\footnote{Gravitational block formulae have also been
obtained in a higher-dimensional context in \cite{Martelli:2023oqk,Colombo:2023fhu} using the equivariant volume of symplectic toric orbfolds. However, spinor bilinears and the associated canonical procedure for going off-shell 
that are utilised here and in \cite{BenettiGenolini:2023kxp,BenettiGenolini:2023yfe,BenettiGenolini:2023ndb}
were not considered in \cite{Martelli:2023oqk,Colombo:2023fhu}.} and very different techniques for spherical horizons in \cite{Hosseini:2019use,Gauntlett:2019roi} and spindle horizons \cite{Boido:2022iye,Boido:2022mbe}, respectively.

We also extend our $D=4$ analysis to study a particular model involving hypermultiplets. Specifically, we consider
a model that extends the STU model with a complex scalar field, which also arises as a consistent
truncation of $D=11$ supergravity on $S^7$ \cite{Bobev:2018uxk,Suh:2022pkg}. The model has two $AdS_4$ vacua, one dual to the ABJM theory and the other
dual to the mABJM theory \cite{Warner:1983vz,Ahn:2000aq,Corrado:2001nv,Benna:2008zy,Klebanov:2008vq}. 
In this setting the new off-shell entropy function is actually the same as above, but with additional
constraints on the $\newc^I_\pm$ and fluxes $\Pnew^I$ associated with the fact that the charged fields in the hypermultiplet
break some of the gauge symmetries. In particular, we recover the known off-shell entropy function
for $AdS_2\times S^2$ black hole horizons considered in \cite{Bobev:2018uxk}. For spindle horizons the 
new off-shell entropy function gives on-shell results consistent with those found for \cite{Suh:2022pkg}, which were obtained by a direct analysis of the BPS equations. Here we will also provide examples with properly quantized magnetic flux.

Having analysed gauged supergravity, we can obtain results for 
$\mathcal{N}=2$, $D=4$ ungauged supergravity by simply turning off
the gauge parameters. We construct off-shell entropy functions that immediately eliminate the possibility of
having spindle horizons, leaving just the possibility of $AdS_2\times S^2$ near horizon geometry. 
We show how the entropy function for the standard attractor mechanism can then be recovered.

The above discussion focused on $\mathcal{N}=2$, $D=4$ gauged supergravity, but 
an analogous story unfolds for $D=5$ gauged supergravities coupled to vector and hypermultiplets.
In this case, associated with magnetically charged black strings 
(or rings) in $AdS_5$ with a horizon\footnote{Black rings of ungauged supergravity have a near horizon limit
that is locally isometric to $AdS_3\times S^2$ \cite{Elvang:2004rt,Gauntlett:2004qy}.
Black rings in $AdS_5$ are not known, but if they exist our results can be utilized.}
 that is (locally) 
$AdS_3\times M_2$,
we derive an off-shell 
central charge function using the equivariant calculus, and this computes the central charge of the $d=2$, $\mathcal{N}=(0,2)$ SCFT
dual to the $AdS_3\times M_2$ solution. In a similar manner to the $D=4$ case, our work generalizes the standard attractor mechanism equations for the case when $M_2$ is an $S^2$ 
and, in addition, it is also applicable to spindle horizons.
Explicitly, for the spindle horizon, we find the off-shell central charge
\begin{equation}\label{enfnsspndle5d}
	c=  -\frac{3\pi}{2G_5} \frac{1}{b_0}\left[{\mc{F}(\newc^I_+) - 
	\mc{F}(\newc^I_-})\right]\, ,
\end{equation}
where $\mc{F}$ is the real prepotential of the $D=5$  gauged supergravity. The $\newc_\pm^I$ now satisfy the constraints $\xi_I\newc^I_+=2-\frac{b_0}{n_+}P_+$ and $\xi_I\newc^I_-=2+\frac{b_0}{n_-}P_-$, where the constants $\xi_I\in\mathbb{R}$ determine the FI gauging of the theory, and $P_\pm=1$ or $P_\pm=-1$ determine the 
chirality of the Killing spinor at the poles of the spindle with $\sigma=P_+/P_-$ then
labelling twist and anti-twist.  
The magnetic charges are given by $\Pnew^I=-\frac{1}{b_0}(\newc_+^I-\newc_-^I)$. For $S^2$ horizons we find
 \begin{equation}\label{enfnssphere5d}
c = \frac{3\pi}{2G_5}  \Pnew^I \frac{\partial \ }{\partial \newc^I} \mc{F}(\newc^I) \, ,
\end{equation}
with the $\newc^I$ and magnetic charges constrained via $\xi_I\newc^I=2$ and $\xi_I \Pnew^I=2P_+$, in agreement with \cite{Cacciatori:2003kv}.

For the special case of the $D=5$ STU model we make contact with known results 
for $S^2$ \cite{Cacciatori:2003kv} and the spindle \cite{Ferrero:2020laf,Ferrero:2021etw}. Furthermore, after uplifting on $S^5$ to type IIB we make contact with the off-shell central charge computed
in GK geometry for $AdS_3\times M_7$ solutions of type IIB with $M_7$ an $S^5$ bundle over $M_2$, for spheres
\cite{Gauntlett:2018dpc,Gauntlett:2019pqg} and spindles \cite{Boido:2022mbe}.

We also extend our $D=5$ analysis to study a model involving hypermultiplets, finding results consistent with \cite{Bobev:2014jva} for $S^2$ horizons. For spindle horizons we obtain an off-shell central charge that
after extremization gives rise to the on-shell central charge found in \cite{Arav:2022lzo} obtained just by manipulating the BPS equations.

Finally, we also obtain results for $D=5$ ungauged supergravity by simply turning off
the gauge parameters. We construct off-shell entropy functions that immediately eliminate the possibility of
having spindle horizons, leaving just the possibility of $AdS_3\times S^2$ near horizon geometry. 
We show how the entropy function for the standard attractor mechanism can then be recovered.

The plan of the rest of paper is 
as follows. Since the analysis is slightly simpler, we first discuss the five-dimensional examples before 
turning to those in four dimensions. In section \ref{sec:d5sugramodelvectors} we discuss $D=5$ gauged supergravity, coupled to an arbitrary number of Abelian vector multiplets, including the $D=5$ STU model as a special case.
Section~\ref{sec:d5sugramodelhypers} analyses a model that extends the $D=5$ STU model with a hypermultiplet, and section \ref{sec:d5ungauged} summarizes the results obtained for ungauged supergravity. 
We then move to the parallel analysis in four dimensions, discussing $\mc{N}=2$, $D=4$ gauged supergravity coupled to Abelian vector multiplets in section~\ref{sec:D4SUGRA}, including the $D=4$ STU model as an example, and then the STU model with a hypermultiplet in section~\ref{sec:sugramodel}. Section~\ref{sec:D4UngaugedSUGRA} discusses $\mc{N}=2$, $D=4$ ungauged supergravity and
we conclude with some discussion in section~\ref{sec:Discussion}. The details 
of the construction of the equivariantly closed forms in terms of spinor bilinears are presented in the appendices \ref{app:5d} and \ref{app:4dBilinears}, in $D=5$ and $D=4$, respectively.

\section{\texorpdfstring{$D=5$, $\mathcal{N}=2$}{D=5 N=2} gauged supergravity}\label{sec:d5sugramodelvectors}

We begin by considering $\mathcal{N}=2$ gauged supergravity in $D=5$ dimensions, coupled to an arbitrary number $n$ of Abelian vector multiplets. 
This theory was originally constructed in  \cite{Gunaydin:1984ak}, 
but  we will largely follow the conventions of 
 \cite{Behrndt:1998jd, Behrndt:1998ns, Chamseddine:1999xk}.

The bosonic part of the action is given by
\begin{align}
    S_5 = \frac{1}{16\pi G_5} \int_{M_5}&\left( R_5-\mathcal{V}-\mathcal{G}_{ij}\del_\mu\varphi^i\del^\mu\varphi^j-\frac{1}{2}G_{IJ}{F}_{\mu\nu}^I{F}^{J\mu\nu}\right. \nn \\
\label{5daction}
    &\left.\ \ +\frac{1}{24}C_{IJK}\varepsilon^{\mu\nu\rho\sigma\lambda}{F}_{\mu\nu}^I{F}_{\rho\sigma}^J{A}_\lambda^K\right)\vol_5\, .
\end{align}
Here we use signature $(-1,1,\ldots,1)$, with spacetime indices $\mu,\nu=0,\ldots,4$,  $R_5$ denotes the Ricci scalar, $\vol_5$ is the Lorentzian volume form, and $G_5$ is the Newton constant. 
Recall that minimal $D=5$, $\mathcal{N}=2$ gauged supergravity contains a graviphoton 
${A}$, which in \eqref{5daction} has been combined with the 
$n$ additional vector multiplet gauge fields into ${A}^I$ with field 
strengths ${F}^I=\diff {A}^I$, where 
$I=0,1,\ldots,n$. The $n$ real scalars $\varphi^i$, $i=1,\ldots,n$, have 
likewise been repackaged into $X^I=X^I(\varphi^i)$, where the latter 
satisfy the constraint 
\begin{equation}\label{constraint}
    \mathcal{F}(X^I)\equiv \frac{1}{3!}C_{IJK}X^IX^JX^K=1\, ,
\end{equation}
where $\mathcal{F}$ is the prepotential with
$C_{IJK}$ a totally symmetric constant tensor, which also 
specifies the Chern-Simons coupling in \eqref{5daction}. 

The metrics for the kinetic terms for the gauge fields and scalars  are 
respectively
\begin{equation}\label{metrics}
    G_{IJ}= -\frac{1}{2}\del_I\del_J\log \mathcal{F}\, \Big|_{\mathcal{F}=1}\,, \quad
    \mathcal{G}_{ij}=\del_i X^I \del_j X^J G_{IJ}\, \Big|_{\mathcal{F}=1}\, ,
\end{equation}
where we use the notation $\del_I\equiv \del_{X^I}$, $\del_i\equiv\del_{\varphi^i}$.
The scalar potential is
\begin{equation}\label{V5}
    \mathcal{V} \equiv \xi_I \xi_J\left(\mathcal{G}^{ij}\del_iX^I\del_jX^J-\frac{4}{3}X^I X^J\right)\, ,
\end{equation}
where $\xi_I$ are constant Fayet--Iliopoulos (FI) gauging parameters. 
This  vector multiplet geometry is known as \emph{very special geometry} \cite{deWit:1992cr}, and the following definition and relations will prove useful for what follows:
\begin{equation}\label{GIJ2/3}
    X_I\equiv \frac{2}{3}G_{IJ}X^J\, , \quad \del_{i}X_I=-\frac{2}{3}G_{IJ}\del_{i}X^J\,, 
\end{equation}
together with the expression
\begin{equation}\label{GIJ}
  G^{IJ}=\mathcal{G}^{ij}\del_i X^I\del_j X^J+\frac{2}{3}X^I X^J  \, ,
\end{equation}
for the inverse of $G_{IJ}$. In particular note $X^IX_I=1$ follows from the constraint 
\eqref{constraint}. 

A solution is supersymmetric if there is a non-trivial solution 
$\epsilon$ to the Killing spinor and gaugino equations: 
\begin{align}
      &0=\left[\nabla_\mu-\frac{\ii}{2}\xi_I{A}^I_\mu+\frac{1}{6} W \Gamma_\mu +\frac{\ii}{8}X_I{F}_{\nu\rho}^I(\Gamma_{\mu}^{\ \nu\rho}-4\delta_\mu^\nu\Gamma^\rho)\right]\epsilon\,, \nn \\
\label{5dKSE}
      &0=\left[-\frac{\ii}{2}\mathcal{G}_{ij}\del_\mu\varphi^j\Gamma^\mu+\frac{\ii}{2} \del_i W+\frac{3}{8}\del_i X_I{F}_{\mu\nu}^I\Gamma^{\mu\nu}\right]\epsilon\, .
\end{align}  
Here $\Gamma_\mu$ generate $\mathrm{Cliff}(1,4)$ in an orthonormal frame, $\nabla_\mu$ is the spin connection, and we have defined the superpotential
\begin{align}\label{W5d}
W \equiv \xi_I X^I\, .
\end{align}
Notice that we may then write the potential in \eqref{V5} as
\begin{align}\label{P5d}
\mathcal{V} = \mathcal{G}^{ij}\partial_i W \partial_j W - \frac{4}{3} W^2\, .
\end{align}

If the $D=5$ theory admits a supersymmetric $AdS_5$ solution with radius $R$, then it is dual to a $d=4$, $\mathcal{N}=1$ SCFT 
with central charge given (in the large $N$ limit) by the usual formula
\begin{align}\label{acentchge}
a=\frac{\pi R^3}{8G_5}\,.
\end{align}

\subsection{The \texorpdfstring{$AdS_3$}{AdS3} ansatz}\label{sec:AdS3}

We consider supersymmetric solutions 
of the warped product form
\begin{equation}\label{ansatz5d}
     \dd s_5^2 = \ex^{2\lambda}\left[\dd s^2(AdS_3)+\dd s^2(M_2)\right]\,. 
\end{equation}
Here the metric on $AdS_3$ is taken to have unit radius,
$\lambda$ and the scalar fields are taken to be functions 
on $M_2$, with the gauge fields $A^I$ taken to be (pull-backs of) gauge fields on $M_2$. 

Inserting this ansatz  into the $D=5$ equations of motion gives rise to $D=2$ equations of motion. The latter
can be obtained by varying a two-dimensional action $S_2$ for the fields on $M_2$, which can in turn be obtained by 
substituting the ansatz into the $D=5$ action \eqref{5daction} and writing $S_5=\frac{\Vol(AdS_3)}{16\pi G_5}S_2$. Specifically,
we have
\begin{align}
\label{eq:5d_2dEffective}
    S_2 & =\int_{M_2}\bigg[\ex^{3\lambda}\left(R-6+12(\nabla\lambda)^2-\mathcal{G}_{ij}\del_\mu\varphi^i\del^\mu\varphi^j\right)-\ex^{5\lambda}\mathcal{V}\nonumber\\
& \qquad \qquad -\frac{1}{2}\ex^{\lambda}G_{IJ}F_{\mu\nu}^I F^{J \mu\nu }\bigg]\vol\,,
\end{align}
where $R$, $\nabla$ and $\vol$ are respectively the Ricci scalar, 
Levi--Civita connection and volume form on $M_2$. 
The $D=2$ Maxwell equation 
for $A^I$ reads
   \begin{equation}\label{2dMax}
        \dd\left[\ex^\lambda G_{IJ}(* F^J)\right]=0\, \quad  \implies \quad \dd\left[\ex^\lambda G_{IJ}F_{12}^J\right]=0\, ,\quad \forall I=0,\ldots,n \, .
    \end{equation}
Here we write $F^I = F^I_{12}\, \vol$, where $F^I_{12}$ is a function 
on $M_2$. The equation of motion for the warp factor $\lambda$ 
is
    \begin{align}\label{5dlambdaEOM}
        \ex^{3\lambda}\left[R-6+12(\nabla\lambda)^2-\mathcal{G}_{ij}\del_\mu\varphi^i\del^\mu\varphi^j\right] & = \frac{5}{3}\ex^{5\lambda}\mathcal{V}+\frac{1}{6}\ex^{\lambda}G_{IJ}F_{\mu\nu}^IF^{J \mu\nu}\nonumber\\
& \quad +\mbox{total derivative}\, .
\end{align}
The trace of the $D=2$ Einstein equation reads 
\begin{equation}\label{d2treeq}
		\frac{2}{3}\ex^{5\lambda}\mathcal{V}-\frac{1}{3}\ex^{\lambda}G_{IJ}F_{\mu\nu}^I F^{J \mu\nu }=-4\ex^{3\lambda}+\mbox{total derivative}\, .
\end{equation}

Imposing the equation of motion for the warp factor\footnote{Equivalently, this is
the same as imposing the trace of the $D=5$ Einstein equations. 
} \eqref{5dlambdaEOM} then leads 
to the following form of the partially off-shell (POS)\footnote{The acronym also stands for partially on-shell, for the more optimistic reader.} two-dimensional action: 
\begin{equation}
\label{reducedS}
    S_2|_{\text{POS}} =\frac{2}{3}\int_{M_2}\left(\ex^{5\lambda}\mathcal{V}\, \vol-\ex^{\lambda}G_{IJ}F_{12}^IF^{J}\right)\, .
\end{equation}
We next note that if we were to also impose the trace of the $D=2$ Einstein equation \eqref{d2treeq} we would get 
the \emph{on-shell} action
\begin{equation}\label{osactd5cases}
	S_2|_{\text{OS}}=-4 \int_{M_2} \ex^{3\lambda}\vol\, .
\end{equation}
The central charge of the
dual $d=2$, $\mathcal{N}=(0,2)$ SCFT is given by $c=3/2G_3$, where $G_3$ is the $D=3$ Newton constant. By analogy with $c$-extremization in 
field theory \cite{Benini:2012cz} we then introduce the  ``trial''
central charge function 
\begin{align}\label{centchgeform}
c=-\frac{3}{8G_5}S_2|_{\text{POS}}=-\frac{3a}{\pi R^3}S_2|_\text{POS}\,.
\end{align}
This has the property that for an on-shell $AdS_3$ solution 
$c$ is the central charge. 
In the last equality we have used \eqref{acentchge} to express the result for $c$ in terms of the central charge $a$ 
of the $D=4$ SCFT (when the $D=5$ model admits an $AdS_5$ vacuum).

Our goal is to compute the central charge of the $d=2$ SCFT for any $AdS_3$ solution without having an explicit solution, 
but instead using localization. To do this we will implement localization using 
the partially off-shell action $ S_2|_{\text{POS}}$ and then carry out a final residual extremization
in order to get the on-shell action and hence central charge $c$.

\subsection{Equivariantly closed forms}\label{sec:5dequivariant}

With the ansatz \eqref{ansatz5d}, 
the Killing spinor $\epsilon$ that solves \eqref{5dKSE} may be decomposed 
as 
\begin{align}\label{epsilon3plus2}
\epsilon = \vartheta\otimes \ex^{\lambda/2}\zeta\, .
\end{align}
Here $\vartheta$ is a Killing spinor on $AdS_3$, 
$\zeta$ is a spinor on $M_2$ and the warp factor is included for convenience. 
The Killing spinor equation \eqref{5d2dKSE} can then be reduced, leading to 
a set of differential and algebraic equations that $\zeta$ must satisfy, as given in
appendix~\ref{app:5d}. 

It is next convenient to introduce the following real bilinears in $\zeta$:
\begin{equation}\label{5dbilinears}
    S = \zeta^\dagger\zeta\,, \quad P = \zeta^\dagger\gamma_3\zeta\,, \quad K = \zeta^\dagger\gamma_{(1)}\zeta\,, \quad \xi^\flat = -\ii\zeta^\dagger\gamma_{(1)}\gamma_3\zeta\,.
\end{equation}
Here we denote $\gamma_{(n)}=\frac{1}{n!}\gamma_{\mu_1\cdots \mu_n}\diff x^{\mu_1}\wedge\cdots\wedge \diff x^{\mu_n}$, and 
$\gamma_3= -\ii \gamma_1\gamma_2$ is the chirality operator on $M_2$. 
These bilinears satisfy a set of algebraic relations and differential conditions that follow from the
reduced Killing spinor equations \eqref{5d2dKSE}. Importantly, one deduces
that the vector field $\xi$ dual to the one-form $\xi^\flat$ is 
a Killing vector on $M_2$ and we also have 
\begin{align}\label{dxiflattext5d}
\dd \xi^\flat & = -2\Big(2+PS^{-1}\ex^{\lambda}W\Big)P\, \vol\,.
\end{align} 

In appendix \ref{app:5d} it is also shown that the multi-forms 
\begin{align}\label{closedPhi}
    \Phi^{F^I}& \equiv F^I-X^I \ex^\lambda P \, ,
   \end{align}
as well as
\begin{align}
 \Phi^\vol& \equiv \ex^{5\lambda}\mathcal{V}\, \vol-\ex^{4\lambda}W S\, ,
\end{align}
are both equivariantly closed under $\diff_\xi = \diff - \xi\hook\, $. 
Using the two-dimensional Maxwell equation \eqref{2dMax}, we can immediately construct another equivariantly closed form,
\begin{align}\label{phiSexp5d}
\Phi^S&\equiv \frac{2}{3}\left(\Phi^\vol-\ex^\lambda G_{IJ}F_{12}^I\Phi^{F^J}\right)\nn\\
&=\frac{2}{3}\left(\ex^{5\lambda}\mathcal{V}\, \vol-\ex^{\lambda}G_{IJ}F_{12}^IF^{J}
-\ex^{4\lambda}WS+\ex^{2\lambda} G_{IJ}F_{12}^IX^J P\right)\,,
\end{align}
and we notice the top-form is precisely that appearing in the partially off-shell two-dimensional
action \eqref{reducedS}.

It is also shown in appendix \ref{app:5d} that the scalar bilinear $S$ is constant, and 
without loss of generality we normalize $S\equiv 1$.

\subsection{Localization}
\label{sec:5dLocalization}

From the definitions \eqref{5dbilinears} one can show 
that $\|\xi\|^2=\|K\|^2=S^2-P^2$. 
At a point $p$ where $\xi|_p\, =0$ it immediately follows that 
$P|_p \, =\pm S|_p \, = \pm 1$, where 
this sign is the chirality of the spinor $\zeta$ 
at the point $p$.   
Provided 
$\xi$ is not identically zero\footnote{If $\xi$ is identically zero then
so is $K$, and $S\equiv P\equiv 1$. One quickly deduces
from the equations in appendix \ref{app:5d}
that $\lambda$ and $X^I$ are constant on $M_2$, 
and localization plays no immediate role in this case. 
In particular this is the case for the topological 
twist solutions in which $M_2$ is a constant curvature 
Riemann surface. 
However, when $M_2=S^2$ we are still able to obtain a non-trivial result for the off-shell
action using localization by taking a limit 
of our fixed point formula, and also can recover the result 
that $\xi$ is identically zero on $S^2$ --  see
section \ref{sec:limit}. \label{foot}}, 
near the point $p$ we may model $M_2$ as 
$\R^2/\Z_{d_p}$, where we allow for potential 
orbifold singularities, and 
in these local coordinates write $\xi = \epsilon_p 
\partial_{\phi_p}$, where $\phi_p$ is a 
polar coordinate on $\R^2/\Z_{d_p}$ with period $2\pi/d_p$. 
Notice that the orientation here is determined by the 
global orientation on $M_2$ induced by the volume form. 
We refer to $\epsilon_p$ as the \emph{weight} of 
$\xi$ at the fixed point $p$. 

With this notation to hand, we may apply the 
Berline--Vergne--Atiyah--Bott fixed point formula 
to evaluate integrals of the 
equivariantly closed forms in \eqref{closedPhi} and \eqref{phiSexp5d} over $M_2$. 
We have the magnetic charges\footnote{
In general there is no reason to quantize the magnetic charges. For models that can be uplifted to string/M-theory, such as the STU model discussed
in section \ref{sec:5dSTU}, the gauge fields are associated with $U(1)$ bundles and there is a quantization condition. \label{footnote:quantization_5d}}
\begin{align}\label{fluxp}
\Pnew^I \equiv \int_{M_2}\frac{F^I}{2\pi} = \int_{M_2}\frac{\Phi^{F^I}}{2\pi}  = -\frac{1}{2\pi}\sum_{\mathrm{fixed\, } p}
\frac{1}{d_p}\frac{2\pi}{\epsilon_p}(X^I\ex^\lambda P)\bigg|_p\, .
\end{align}
We may then similarly localize the action \eqref{reducedS} using \eqref{phiSexp5d}
\begin{align}\label{actionp}
    S_2|_{\text{POS}}
    =\int_{M_2} \Phi^S\nonumber 
    &=-\frac{2}{3}\sum_{\mathrm{fixed\, } p}\frac{1}{d_p}\frac{2\pi}{\epsilon_p}\Big(\ex^{4\lambda}WS-\ex^{2\lambda} G_{IJ}F_{12}^IX^J P\Big)\Big|_p\nonumber\\
     &=2\sum_{\mathrm{fixed\, } p}\frac{1}{d_p}\frac{2\pi}{\epsilon_p}\ex^{3\lambda}P\, \Big|_p\, ,
\end{align}
where  in the 
last equality we have used the algebraic equation \eqref{5dalg}, 
and then used that $P^2|_p\, = S^2|_p\, = 1$. 
The final constraint comes from the equation for $\diff\xi^\flat$ in 
\eqref{dxiflattext5d}. Since at a fixed point 
$\diff\xi^\flat\, |_p = 2\epsilon_p \vol$, in terms of the  
weight $\epsilon_p$ of $\xi$ at $p$, we deduce that  
\begin{align}\label{superp}
\ex^\lambda  WP|_p \, = -(2 +  \epsilon_p P_p)\, ,
\end{align}
where recall that $P_p=\pm 1$, with the sign according to the spinor chirality at $p$. 

In order to  evaluate these expressions further, it is helpful to first introduce the 
new scalar fields
\begin{align}
\label{eq:IntroductioncI}
    \newc^I \equiv -X^I\ex^\lambda P\, \quad \implies \quad \Phi^{F^I} = 
F^I + \newc^I\, .
\end{align}
Using the constraint $\mathcal{F}(X^I)=1$ in \eqref{constraint}, at 
a fixed point $p$ we may then compute
\begin{align}
\mathcal{F}(\newc^I) |_p\,  = \frac{1}{3!}C_{IJK} \newc^I \newc^J\newc^K|_p \, = -\ex^{3\lambda}P|_p\, ,
\end{align}
where we have again used $P^2|_p\, =1$. 
We may then rewrite \eqref{actionp}, \eqref{fluxp}, \eqref{superp} as 
\begin{align}\label{cIpformulas}
 S_2|_{\text{POS}} =-2 \sum_{\mathrm{fixed\, } p}\frac{2\pi}{d_p \epsilon_p}\mathcal{F}(\newc^I_p)\, , \quad \Pnew^I = \sum_{\mathrm{fixed\, } p}\frac{1}{d_p \epsilon_p}\newc^I_p\, , \quad \xi_I \newc^I_p = 2 + \epsilon_p P_p\, ,
\end{align}
respectively, where $\newc^I_p \equiv \newc^I|_p$. This makes manifest 
that the action may be computed knowing only the 
values of the scalar fields $\newc^I_p$ at the fixed points, where these 
are constrained to satisfy the equations in \eqref{cIpformulas}, involving 
the magnetic charges $\Pnew^I$ and the data $d_p, \epsilon_p, P_p$ at the fixed points. 

If $M_2$ is compact without boundary, then the only such space admitting 
a $U(1)$ isometry with fixed points is topologically a spindle 
$M_2\cong \mathbb{WCP}^1_{[n_+,n_-]}$.\footnote{One could 
in principle have $M_2=T^2$, but in this case any non-trivial Killing vector 
is fixed point free, and our formulae immediately imply 
that the magnetic charges $P^I$ are zero, and the action is zero, 
ruling out such solutions. Note that this argument 
is not in contradiction with the existence of the $T^2$ solutions constructed in 
\cite{Almuhairi:2011ws} -- it only says that such solutions do not have a non-zero 
R-symmetry Killing vector along the $T^2$. Unlike the case of $S^2$ discussed in footnote \ref{foot} and section
\ref{sec:limit}, we cannot deduce more about the $T^2$ case using our approach.} Here there are two 
fixed points, which we label by $p=\pm$,
with $d_\pm = n_\pm$ and we take $n_\pm\ge 1$. We may write the Killing vector field globally as
\begin{align}
\xi = b_0 \partial_{\varphi}\, ,
\end{align}
where $\varphi$ is an azimuthal coordinate on the spindle with period $2\pi$. 
Then without loss of generality we write $\epsilon_+ = -b_0/n_+$, $\epsilon_- = b_0/n_-$, but leave the signs of $P_\pm$ arbitrary, where recall $|P_\pm|=1$. 
The constraint equations in \eqref{cIpformulas} now read  
\begin{align}\label{cIpm}
\Pnew^I = -\frac{1}{b_0}(\newc_+^I - \newc_-^I)\, , \quad  \xi_I\newc^I_+ = 2-\frac{b_0}{n_+}P_+\, , \quad 
 \xi_I \newc^I_- = 2+\frac{b_0}{n_-}P_-\, ,
\end{align}
and combining these equations one deduces 
\begin{align}\label{twistme}
    \xi_I \Pnew^I=\frac{P_+}{n_+}+\frac{P_-}{n_-}=\frac{n_-P_++ n_+P_-}{n_-n_+}\,.
\end{align}
Since from \eqref{5dKSE} the Killing spinor $\epsilon$ is charged 
under precisely the combination $\xi_IA^I/2$, equation \eqref{twistme}
gives  the magnetic flux of the R-symmetry gauge field through the spindle $M_2$. The form  of the right hand side of \eqref{twistme} was proven 
very generally in \cite{Ferrero:2021etw}. Introducing 
$\sigma\equiv  P_+/P_-$, the cases $\sigma=1$, $\sigma=-1$ correspond 
to the spinor chirality at the two poles being either the same or opposite, respectively;  
in turn, these were called the  twist 
and anti-twist in \cite{Ferrero:2021etw}, respectively.\footnote{More 
precisely, from equation (3.21) of \cite{Ferrero:2021etw} we see that 
$P_+$, $P_-$ are equivalent to the variables $\eta_1,\eta_2$ in \cite{Ferrero:2021etw}, 
which are also $\pm1$.}  
The off-shell action itself takes the compact form
\begin{align}\label{actioncIpm}
S_2|_{\text{POS}} = \frac{4\pi}{b_0}\left[\mathcal{F}(\newc_+^I) - \mathcal{F}(\newc_-^I)\right]\, ,
\end{align}
and using \eqref{centchgeform} gives our final expression for the off-shell central charge.

To obtain the on-shell central charge, we need to extremize over the undetermined data and
we can proceed as follows.
First, the $\newc_-^I$ 
are determined by the $\newc_+^I$ and magnetic charges $\Pnew^I$ via the first equation
in \eqref{cIpm}:
\begin{align}\label{cIm}
\newc_-^I=\newc_+^I + b_0\, \Pnew^I\, .
\end{align}
 The magnetic charges $\Pnew^I$ are constrained to satisfy 
the twist/anti-twist condition \eqref{twistme}, while the $\newc_+^I$ 
satisfy a single constraint in \eqref{cIpm} (which is a constraint 
on the superpotential). This leaves $n$ degrees of freedom in the $\newc_+^I$, 
and $b_0$ 
and  one then varies the action in \eqref{actioncIpm} over these degrees 
of freedom to obtain the final on-shell result. We will carry out this 
extremization explicitly for the STU model further below.

\subsection{The \texorpdfstring{$b_0\rightarrow 0$}{b0 to 0}  limit}\label{sec:limit}

Notice that substituting \eqref{cIm} into \eqref{actioncIpm} allows us 
to compute the limit 
\begin{align}\label{cIquad}
\lim_{b_0\rightarrow 0} S_2|_{\text{POS}} =-4\pi \Pnew^I\frac{\partial}{\partial \newc^I}
 \mathcal{F}(\newc^I) =-{2\pi}C_{IJK}\Pnew^I \newc^J \newc^K\, .
\end{align}
Here in the limit $\newc_-^I=\newc_+^I=\newc^I$ is constant on $M_2$. 
This is of course consistent with our comments in footnote \ref{foot}. In addition, from
\eqref{cIpm} we have that the $\newc^I$ are subject to the constraint $\xi_I\newc^I=2$.
Our general procedure implies that for an on-shell solution one should extremize 
the quadratic function of the $\newc^I$ in \eqref{cIquad}, subject
to the constraint $\xi_I\newc^I=2$. In particular setting 
$n_+=n_-=1$ this recovers the result in 
\cite{Cacciatori:2003kv} for $S^2$ solutions with a topological twist, where 
the magnetic charges $\Pnew^I$ are constrained via \eqref{twistme} with $\sigma=+1$ to satisfy
$\xi_I\Pnew^I=2P_+$.

It is remarkable that we have obtained these non-trivial results for $AdS_3\times S^2$ solutions with a topological twist which have $b_0=0$, 
not by taking $b_0=0$ at the beginning but instead assuming $b_0\ne0$ and then taking the limit $b_0\to 0$.

\subsection{STU model}\label{sec:5dSTU}

The STU model is a consistent truncation of type IIB supergravity on
 $S^5$, with the $U(1)^3=SO(2)^3\subset SO(6)$ isometry gauged in $D=5$.
In the conventions used in \cite{Ferrero:2021etw}, this may be obtained
 from the general $D=5$, $\mathcal{N}=2$ 
gauged supergravity theory by setting $n=2$, 
$\xi_I=(1,1,1)$ and where the only 
non-zero component of $C_{IJK}$  (up to permutations) is $C_{123}=1$. 
This model then has an $AdS_5$ vacuum which has unit radius, {\it i.e.} $R=1$ in \eqref{acentchge}, 
which is dual to $\mathcal{N}=4$ SYM theory with $a$ central charge given by $a=N^2/4$. The model
is explicitly written down in section \ref{sec:d5sugramodelhypers} (where one should set $\vpvar=0$ there).
As discussed in section \ref{sec:UpliftAdS3}, the quantization condition on the magnetic fluxes
that enables one to uplift the solutions to $D=10$ on $S^5$ is such that if we write
$\Pnew^I=p_I/(n_+n_-)$ we have $p_I\in\mathbb{Z}$.

Our result for the off-shell central charge  function \eqref{centchgeform} in this set-up reads 
\begin{align}\label{cstuoffs}
c=-\frac{3N^2}{b_0}\left(\newc_+^1\newc_+^2\newc_+^3-\newc_-^1\newc_-^2\newc_-^3\right)\,,
\end{align}
with 
\begin{align}\label{cIpm2}
\Pnew^I = -\frac{1}{b_0}(\newc_+^I - \newc_-^I)\, , \quad \sum_{I=1}^3\newc^I_+ = 2- \frac{b_0}{n_+}P_+\, , \quad 
\sum_{I=1}^3 \newc^I_- = 2+\frac{b_0}{n_-}P_-\, ,
\end{align}
and the sum of the magnetic charges constrained to obey $\sum_{I=1}^3 \Pnew^I=\frac{n_-P_++n_+P_-}{n_-n_+}$,
or equivalently, $p_1+p_2+p_3=n_-P_++ n_+P_-$.

It is straightforward to extremize over the independent variables $b_0,\newc^1_+,\newc^2_+$, say, and we find the following on-shell results.  
For $c$ and $b_0$ we now get 
\begin{align}
c&=\frac{6p_1 p_2 p_3}{n_+n_-\svar}N^2\,,\qquad b_0=\frac{2n_+n_-(n_-P_+-n_- P_+)}{\svar}\,,
\end{align}
where
\begin{align}
s\equiv n_+^2+n_-^2-(p_1^2+p_2^2+p_3^2)\,.
\end{align}
For the variables $\newc_\pm^I$ evaluated at the poles we find 
\begin{align}
\newc^I_\pm=\frac{2p_I(n_\pm P_\mp-p_I)}{\svar}\,.
\end{align}

Recall that we took $n_\pm\ge 1$ and $|P_\pm|=1$, so
from \eqref{eq:IntroductioncI} we have $P_+\newc^I_+<0$ and $P_- \newc^I_-<0$. From \eqref{osactd5cases} we also have
$c>0$. We find that for the twist case where $\sigma=P_+/P_-=+1$ 
there are no solutions with $P_+=P_-=1$, but for $P_+=P_-=-1$ 
the above inequalities reduce to 
precisely two of the $p_I$ being positive, which since 
$p_1+p_2+p_3<0$ in this case then implies that the third 
$p_I$ is negative. 
For the anti-twist case with $\sigma=P_+/P_-=-1$ we instead find that 
the inequalities reduce simply to $p_I>0$. 

These results are in precise  agreement with those 
 found for the explicit supergravity solutions given in \cite{Ferrero:2021etw}. 
Notice that we are then able to obtain expressions for various other quantities, without
having to solve any supergravity equations directly. For example, one can compute 
the warp factors at the poles of the spindle:
\begin{align}\label{e3lforstufive}
\ex^{3\lambda} P |_{\pm}\,  = \frac{8\prod_{I=1}^3 p_I (p_I - n_\mp P_\pm)}{s^3}\, .
\end{align}

\subsection{Uplifting to \texorpdfstring{$AdS_3\times Y_7$}{AdS3 x Y7} solutions}
\label{sec:UpliftAdS3}

The STU solutions of the previous subsection may be 
uplifted on $S^5$ to solutions of type IIB supergravity, where 
the anti-twist class were first constructed and uplifted in 
\cite{Hosseini:2021fge, Boido:2021szx}. The metric on the $S^5$ 
internal space takes the fibred form
\begin{align}\label{fibredS5}
\diff s^2_{S^5} = \sum_{I=1}^3 (X^I)^{-1}\left[\diff \mu_I^2+ \mu_I^2 (\diff \psi_I + A^I)^2\right]\, .
\end{align}
Here $(\mu_I,\psi_I)$ form a system of polar coordinates on 
$S^5\subset\R^2\oplus\R^2\oplus\R^2$, where correspondingly 
$\sum_{I=1}^3\mu_I^2=1$ and the angular coordinates $\psi_I$ each have period $2\pi$. 
The solutions of type IIB then take the form $AdS_3\times Y_7$, 
where $Y_7$ is the total space of a fibration
\begin{align}\label{Y7}
S^5\, \hookrightarrow\, Y_7 \, \rightarrow\, M_2\, .
\end{align}
Here $M_2\cong \mathbb{WCP}^1_{[n_+,n_-]}$ is a spindle, and 
from \eqref{fibredS5} one can see that $S^5$ is fibred as an associated 
bundle for the action of $U(1)^3$ on $S^5\subset\R^2\oplus\R^2\oplus\R^2$, 
where each $U(1)$ is fibred with connection one-form given by the 
$D=5$ supergravity gauge field $A^I$, $I=1,2,3$. 
The magnetic charges $\Pnew^I=p_I/(n_+n_-)$, with $p_I\in\Z$, 
are then precisely Chern numbers for this fibration, with 
their quantization condition being equivalent to \eqref{Y7} being a well-defined orbifold fibration \cite{Ferrero:2021etw}.
The solutions have only the RR five-form flux turned on in type IIB, and 
may be interpreted as the near-horizon limits 
of $N$ D3-branes wrapped over the spindle $M_2$.

On the other hand, supersymmetric $AdS_3\times Y_7$ solutions 
of type IIB supergravity with only RR five-form flux were first studied in 
\cite{Kim:2005ez}. The results of \cite{Gauntlett:2007ts} show that the existence of a solution to the corresponding 
Killing spinor equation on $Y_7$
is equivalent to imposing certain geometric conditions on $Y_7$. 
The resulting geometric structure was 
dubbed ``GK geometry''  in  \cite{Couzens:2018wnk}. 
 In particular 
$Y_7$ is equipped 
with an R-symmetry Killing vector field $\xi$ that is constructed as a bilinear 
in the Killing spinor.  
In \cite{Couzens:2018wnk} an {\it a priori} 
different  approach to defining an off-shell 
central charge function was introduced in this setting. 
There one first introduces an action for the supergravity fields on $Y_7$, where
solutions to the equations of motion extremize this action; one 
then imposes that the fields are such that a solution 
to the Killing spinor equation exists, which as mentioned 
is equivalent to imposing a certain geometric structure 
(an ``off-shell GK geometry'', in the terminology of 
\cite{Couzens:2018wnk}). The resulting 
supersymmetric action ({\it i.e.} the action with supersymmetry imposed) 
was shown to be a function of a trial R-symmetry vector $\xi$, 
where solutions to the equations of motion extremize this 
function over the choice of $\xi$. Moreover, at the critical point 
this action is proportional to the central charge of the $AdS_3$ 
solution.

We refer the reader to \cite{Couzens:2018wnk, Boido:2022mbe}
for further details, here just quoting the result when the internal 
space $Y_7$ takes the fibred form \eqref{Y7}. From section 5.1 of 
 \cite{Boido:2022mbe} we can write down the supersymmetric action/central 
charge function in our conventions\footnote{Specifically, 
$b_0^{\mathrm{there}}=-b_0$, $m_\pm^{\mathrm{there}}=n_\pm$. Although the variables $R^I_\pm$ 
were already introduced in  \cite{Boido:2022mbe}, they are related to the 
variables $b_I^\pm$ used more extensively in that paper  via 
$R^1_\pm = b_1^\pm - b_2^\pm -b_3^\pm$, $R^2_\pm=b_2^\pm$, 
$R^3_\pm = b_3^\pm$. This simply corresponds 
to a different choice of basis for the $U(1)^3$ action on $S^5$ in 
which the Killing spinor has charge $b_1/2=1$ under the first $U(1)$ 
direction and is uncharged under the other two. We refer to 
\cite{Boido:2022mbe} for further details of these variables. }
\begin{align}\label{Ztrial}
\mathscr{Z}=-\frac{3N^2}{b_0}\left(R^1_+R^2_+ R^3_+ - R^1_- R^2_- R^3_-\right)\, . 
\end{align}
Here geometrically the $R^I_\pm$ parametrize the R-symmetry Killing 
vector on the copies of $S^5/\Z_{n_\pm}$ that lie over the 
 poles of the spindle:
\begin{align}
\xi|_{\pm} \, = \sum_{I=1}^3 R_\pm^I \partial_{\psi_I}\, .
\end{align}
These obey the constraints (see section 7.4 of \cite{Boido:2022mbe})
\begin{align}\label{Rcons}
R^I_+ - R^I_- = -b_0\frac{M^I}{N} = -b_0 \Pnew^I\, , \quad 
\sum_{I=1}^3 R^I_+ = 2 + \frac{b_0}{n_+}\, , \quad 
\sum_{I=1}^3 R^I_- = 2 - \frac{b_0}{\sigma n_-}\, .
\end{align}
Here in the type IIB solutions the $M^I$ arise as quantized fluxes of the 
RR five-form through certain five-cycles in $Y_7$, but 
in this particular case these can also be straightforwardly 
identified with the Chern numbers $M^I/N=\Pnew^I=p_I/(n_+n_-)$ 
of the fibration in \eqref{Y7}. In \cite{Boido:2022mbe} it is 
also shown that $N$ is divisible by $n_+n_-$, so that the $M^I\in\Z$ 
are correctly Dirac quantized fluxes.
The constraints \eqref{Rcons} precisely agree with \eqref{cIpm2} provided
we set $P_+=-1$ and recall $\sigma=P_+/P_-$, 
with $\sigma=\pm1$ being the twist and anti-twist respectively,
 and identify
\begin{align}\label{Rtox}
R^I_\pm = \newc^I_\pm\, .
\end{align}
Moreover, the trial central charge function  \eqref{Ztrial} in GK geometry
precisely matches our trial central charge function \eqref{cstuoffs}, defined 
in $D=5$ STU gauged supergravity.

Another set of variables was also introduced in \cite{Boido:2022mbe}, 
so as to match to both a field theory expression and corresponding 
conjectures for gravitational block formulae in \cite{Hosseini:2021fge, Faedo:2021nub}. 
We write 
\begin{align}
\newc_\pm^I = \phi^I \mp \frac{b_0}{2}\Pnew^I\, .
\end{align}
By virtue of \eqref{cIpm}, \eqref{twistme}, 
the $\phi^I$ variables then satisfy the constraint
\begin{align}
\sum_{I=1}^3 \phi^I = 2 - \frac{b_0}{2}\frac{n_-P_+ - n_+ P_-}{n_+n_-}\, .
\end{align}
The central charge expression \eqref{cstuoffs} then reads 
\begin{align}
\label{eq:5dGaugedSTU_cOff}
c  = 3N^2\left[\Pnew^1\phi^2\phi^3
+\Pnew^2\phi^3\phi^1+ \Pnew^3\phi^1\phi^2+\frac{b_0^2}{4}\Pnew^1\Pnew^2\Pnew^3\right]\, ,
\end{align}
in precise agreement with (5.11) of 
\cite{Boido:2022mbe}.

The agreement of the off-shell functions \eqref{cstuoffs}, \eqref{Ztrial}, 
with the identification of variables \eqref{Rtox}, 
seems to be something of a miracle, since the former is defined purely in $D=5$ 
gauged supergravity, while the latter is defined in terms of the GK geometry 
of the internal space $Y_7$. However, the identification \eqref{Rtox} 
itself is
straightforward to explain geometrically. 
The equivariantly closed forms $\Phi^{F^I}/2\pi$ introduced in 
\eqref{closedPhi} are by construction representatives of the equivariant extensions of the first Chern class of
the complex line bundles $L^I$ on which the $A^I$ are connection one-forms. 
Denoting these equivariant first Chern classes 
$c_1^\xi(L^I)$, at a fixed point $p$ 
we then have the standard property of 
equivariant Chern classes
\begin{equation}
\label{eq:EquivariantChernClasses}
	\left.c_1^\xi(L^I)\right|_{p} = \frac{\epsilon_p(L^I)}{2\pi} \, ,
\end{equation}
where $\epsilon_p(L^I)$ is the  \emph{weight} 
of the action of the vector field $\xi$ on the fixed complex line over the point $p$. 
On the other hand, from \eqref{eq:IntroductioncI} we then 
have
\begin{align}
\epsilon_\pm (L^I) = \newc^I_\pm\, .
\end{align}
The variables $\newc^I_\pm$ are thus precisely weights 
of the lifted R-symmetry vector field on the three fixed complex 
lines $\C\oplus\C\oplus\C=\R^6$ over  each pole of the spindle,
 which is also precisely what the variables $R^I_\pm$ are in GK 
geometry. That is, 
upon uplifting the STU model solution to a IIB solution, the R-symmetry vector $\xi$ acting on $M_2$ is lifted to the R-symmetry vector  acting on $Y_7$. This at least partially
 explains why the formula \eqref{cstuoffs} for the off-shell central charge matches the expression \eqref{Ztrial} taken from \cite{Boido:2022mbe}, in terms of these weights being identified. 

Clearly this argument is not restricted to the STU model: if there is an uplift of a gauged supergravity, then the values of the scalars $\newc^I$ at the fixed points in $M_2$ should always be identified with the weights of the R-symmetry vector in the solution of the uplifted theory.

\section{\texorpdfstring{$D=5$}{D5} STU model with hypermultiplets}\label{sec:d5sugramodelhypers}

We now consider extending the results of the previous section to include hypermutiplets. We
will restrict our considerations to a specific model that extends the STU model, for which $AdS_3\times M_2$
solutions, with $M_2$ a spindle, were analysed in \cite{Arav:2022lzo}.

We start by considering a truncation of  maximal $SO(6)$ gauged supergravity discussed in \cite{Bobev:2010de}, extending \cite{Khavaev:2000gb}. 
One first considers a $\mathbb{Z}_2\times\mathbb{Z}_2\subset SO(6)$ 
invariant sector which gives rise to an $\mathcal{N}=2$ gauged supergravity theory with two vector multiplets and 
4 hypermultliplets. Then one utilizes an additional $\mathbb{Z}_4\subset SO(6)\times SL(2)$ symmetry, as in \cite{Khavaev:2000gb}, 
to further truncate the hypermultiplets. This gives a theory whose bosonic content consists of a metric, three gauge fields 
$A^I$, two real and neutral scalars $\varphi^i$ that live in the $\mathcal{N}=2$ vector multiplets, and four
complex and charged scalar fields $\Zzeta_j$ that are maintained from the hypermultiplets and parametrise
the coset $SU(1, 1)/U(1)$. For our purposes, as in \cite{Arav:2022lzo}, we further truncate to a single complex scalar $\Zzeta$, 
and also set the $D=5$ Chern--Simons term to zero, which is sufficient for the solutions that we study.

In effect we have the STU model coupled to one complex scalar field, $\Zzeta=\vpvar\,  \ex^{\ii\theta}$, that lives in a hypermultiplet and is charged under a certain linear combination of the gauge fields. The bosonic Lagrangian in a mostly plus signature 
is given\footnote{
We have obtained this from \cite{Arav:2022lzo} by taking
$\alpha\to \frac{1}{2\sqrt{6}}\varphi^1$, $\beta\to -\frac{1}{2\sqrt {2}}\varphi^2$,
$\varphi\to\frac{1}{2}\vpvar$, $g_{\mu\nu}\to -g_{\mu\nu}$, $A\to\frac{c_1}{2 }A$,
$g\to 2c_2$, $\gamma^\mu\to \ii c_3\gamma^\mu$
with $c_i=\pm 1$ and $c_1 c_2=-1$, $ c_2 c_3=+1$. We have also redefined $W\to-\frac{1}{2}W$, $\mathcal{P}\to \mathcal{V}/4$.}
 by
\begin{align}\label{d3lagoverall}
\mathcal{L}\, = \, \, 
\frac{1}{16\pi G}\sqrt{-g}&\Big[R-\mathcal{V}-\frac{1}{2}\sum_{i=1}^2(\partial \varphi^i)^2
-\frac{1}{4} \sum_{I=1}^{3}\left(X^{I}\right)^{-2}(F^{I})^{2}
\nn\\
&-\frac{1}{2}(\partial\vpvar)^2-\frac{1}{2}\sinh^2\vpvar (D\theta)^2\Big]\,.
\end{align}
Here $A^{I}$ are three $U(1)$ gauge fields, $I=1,2,3$, with field strengths $F^{I}=\diff A^{I}$. 
The $X^I$ satisfy the constraint $X^{1}X^{2}X^{3}=1$ and are given by
\begin{align}\label{X5d}
X^{1}\, = \, \ex^{-\frac{\varphi^1}{\sqrt{6}}-\frac{\varphi^2}{\sqrt{2}}}\,, \qquad
X^{2}\, = \, \ex^{-\frac{\varphi^1}{\sqrt{6}}+\frac{\varphi^2}{\sqrt{2}}}\,, \qquad
X^{3}\, = \, \ex^{\frac{2\varphi^1}{\sqrt{6}}}\,.
\end{align}
The potential is given by
\begin{align}\label{P5dhpyer}
\mathcal{V}=2\bigg[\sum_{i=1}^2(\partial_{\varphi^i} W)^2+(\partial_\vpvar W)^2\bigg]-\frac{4}{3}W^2\,,
\end{align}
where the superpotential $W$ is given by
\begin{align}
\label{superpottextLS}
W=\sum_{I=1}^3 X^I+\sinh^2\frac{\vpvar}{2}\, (\zeta_IX^I)\,, 
\end{align}
with  $\zeta_I=(1,1,-1)$. In addition the complex scalar is charged with respect to the linear combination of gauge fields given by
$\zeta_IA^I$, and we have
\begin{align}
D\theta\equiv \dd\theta-\zeta_IA^I\,.
\end{align}

For a bosonic solution to preserve supersymmetry, we require 
\begin{align}\label{5d_susy1}
\Big[\nabla_\mu-\frac{\ii}{2}Q_\mu+\frac{1}{6}W\Gamma_\mu
+\frac{\ii}{24}\sum_I(X^I)^{-1}F_{\nu\rho}^{I}(\Gamma_\mu{}^{\nu\rho}-4\delta^\nu_\mu\Gamma^\rho)\Big]\epsilon&=0\,,\nn\\
\Big[\Gamma^\mu{\partial}_\mu\varphi^i-2\partial_{\varphi^i} W+\frac{\ii}{2}\sum_{I=1}^3\partial_{\varphi^i} \left(X^{I}\right)^{-1}\,{F}^{(I)}_{\mu\nu}\Gamma^{\mu\nu}\Big]\epsilon&=0 \,,\nn\\
\Big[\Gamma^\mu{\partial}_\mu\vpvar-2\partial_\vpvar W+2 \ii \partial_\vpvar Q_\mu\Gamma^\mu\Big]\epsilon&=0 \,,
\end{align}
where 
\begin{align}
Q_\mu=\sum_{I=1}^3 A^I_\mu-\sinh^2\frac{\rho}{2}\, D_\mu\theta\,.
\end{align}

The model admits an $AdS_5$ vacuum solution with unit radius and vanishing scalar fields, which is dual to $\mathcal{N}=4$ SYM theory. The $D=5$ Newton constant is given by $\frac{1}{G_5}=\frac{2N^2}{\pi}$ so that $a_{\mathcal{N}=4}=\frac{N^2}{4}$.
There is another $AdS_5$ vacuum with
\begin{align}
\ex^{\frac{\sqrt{3}}{\sqrt{2}}\varphi^1}=2\, ,\qquad \varphi^2=0\, ,\qquad \ex^\vpvar=3\,,
\end{align}
and radius $L_{LS}=3/2^{5/3}$. After uplifting to type IIB supergravity \cite{Pilch:2000ej}
this solution is dual to the $d=4$, $\mathcal{N}=1$ Leigh--Strassler (LS) SCFT \cite{Leigh:1995ep}; the latter 
arises as the IR limit of an RG flow from $\mathcal{N}=4$ SYM theory deformed by a mass deformation and the corresponding holographic solution was found in 
\cite{Freedman:1999gp}. The central charge of the LS SCFT, in the large $N$ limit, is given by $a_{LS}=\frac{27}{32}a_{\mathcal{N}=4}=\frac{27}{128}N^2$.

\subsection{The \texorpdfstring{$AdS_3$}{AdS3} ansatz}

We again consider supersymmetric solutions of the form \eqref{ansatz5d} 
with the warp factor $\lambda$, the scalars and the gauge fields all defined on $M_2$ and $\dd s^2(AdS_3)$ having unit radius. 
We can obtain a $D=2$ action
by substituting this ansatz into the $D=5$ action, and this then gives rise to the correct $D=2$ equations of motion.
Furthermore, after imposing the $D=2$ equations of motion for $\lambda$, or equivalently the trace of the $D=5$ Einstein equations,
we obtain the partially off-shell action ({\it cf.} \eqref{reducedS})
\begin{equation}\label{actionLS}
    S_2 |_{\text{POS}}=\frac{2}{3}\int_{\mathcal{M}^2}\Big[\ex^{5\lambda}\mathcal{V} \, \vol-\frac{1}{2}\sum_{I=1}^3\ex^\lambda(X^I)^{-2}F^I_{12} F^I\Big]\,,
\end{equation}
where $F^I=F^I_{12}\vol$. We may then similarly define 
a trial central charge function
\begin{align}\label{cLScaseitofa}
c=-\frac{3}{\pi}a_{\mathcal{N}=4}\, S_2 |_{\text{POS}}=-\frac{32}{9\pi}a_{LS}\, S_2 |_{\text{POS}}\,, 
\end{align}
which on-shell is the central charge of the $AdS_3$ solution.

\subsection{Equivariantly closed forms and localization}
The ansatz for the Killing spinor is exactly as discussed in section \ref{sec:5dequivariant} and, correspondingly, so are the spinor bilinears. We analyse these further
in the appendix where we show that the one-form dual to the Killing vector again satisfies 
\begin{align}\label{kvcondLS}
\dd\xi^\flat=-2(2+P S^{-1}\ex^\lambda W)P\, \vol\,.
\end{align}
We also use the analysis in appendix \ref{app:5d} to construct the following equivariantly closed forms. 
For the gauge fields we again find that
\begin{equation}
    \Phi^{F^{I}}=F^{I}-X^{I} \ex^\lambda  P\,,
\end{equation}
are equivariantly closed. A new feature in the presence of the hypermultiplet scalars is that
a specific linear combination of the $ \Phi^{F^{I}}$ is equivariantly exact: 
\begin{align}\label{LSdthetacon}
    \dd_\xi D\theta=-\zeta_I\Phi^{F^I}\,.
\end{align}

The polyform that we considered before without hypermultiplets, $\Phi^\vol\equiv \ex^{5\lambda}\mathcal{V}\, \vol- \ex^{4\lambda}W S$,
is no longer equivariantly closed when $\vpvar\ne 0$. However, the following polyform associated with the action \eqref{actionLS},
\begin{align}\label{LSACTFORM}
    \Phi^S=&\ \frac{2}{3}\Big[\Phi^\vol-\frac{1}{2} \sum_{I=1}^3\ex^\lambda (X^{I})^{-2}F^I_{12}
    \Phi^{F^{I}}\Big]\\
    =&\ \frac{2}{3}\bigg[\ex^{5\lambda}\mathcal{V}\, \vol-\frac{1}{2}  \sum_{I=1}^3\ex^\lambda(X^{I})^{-2}F^I_{12}
    {F^{I}}- \ex^{4\lambda}WS
    +\frac{1}{2} \ex^{2\lambda} P\sum_{I=1}^3(X^{I})^{-1}F^I_{12}\bigg]\,,\nn
\end{align}
is equivariantly closed.

The localization then proceeds as in section \ref{sec:5dLocalization} with a few small, but important, changes. We consider $M_2$ to be a spindle
with Killing vector globally defined as $\xi=b_0\partial_\varphi$, with $\Delta\varphi=2\pi$, as before.
It is again convenient to define
\begin{align}\label{cIdefLS}
\newc^I= -\ex^\lambda X^I P\,.
\end{align}
For the fluxes, localization gives
\begin{align}\label{LSBVFLUX}
     \Pnew^I&\equiv \int_{{M}_2} \frac{F^{I}}{2\pi} =-\frac{1}{b_0}(\newc^I_+-\newc^I_-)\,,
     \end{align}
 and for the spindle we have the quantization condition
 \begin{align}\label{LSqnspindle}
     \Pnew^I&=\frac{p_I}{n_-n_+},\qquad p_I\in\mathbb{Z}\,.
     \end{align}
This quantization condition, as in the STU model,
is imposed so that when the solutions are uplifted on $S^5$
we get a well-defined $AdS_3\times M_7$ solution of type IIB supergravity with $M_7$ an $S^5$ fibred over the spindle $M_2$. 
We emphasise that for solutions in which the charged scalar in the 
hypermultiplet is non-zero, we break a $U(1)\subset U(1)^3$ symmetry and, correspondingly, we will lose a Killing vector on
the $S^5$ in the uplifted solution, but this does not alter the fact that we still need to impose a quantization condition on the fluxes to ensure the we have a good $S^5$ fibration over $M_2$, which is a topological constraint. 
 
We can also integrate the new expression \eqref{LSdthetacon} over $M_2$. Since the left hand side is equivariantly exact, the integral vanishes and we immediately deduce that there is a new constraint on the fluxes
\begin{align}\label{brokenLSflux}
\sum_{I=1}^3 {\zeta_I}\Pnew^{I}=0\,,
\end{align}
where $\zeta_I$ were defined in \eqref{superpottextLS}. In addition, we can consider the zero-form component of 
\eqref{LSdthetacon}. By regularity, if $\vpvar|_\pm\ne 0$, we must have $D\theta|_\pm =0$ in order that
the complex scalar is regular at the poles (in the orbifold sense), and hence we deduce the constraints
\begin{align}\label{brokencsLS}
\zeta_I \newc^I_+=\zeta_I \newc^I_-=0\,.
\end{align}
Notice, in particular this implies at the poles $\zeta_IX^I|_\pm=0$ and hence from 
\eqref{superpottextLS}, we find that the value of the superpotential $W$ at the poles only depends
on the scalars in the vector multiplets as in the STU model:
\begin{align}\label{Wexppoles}
W|_\pm =\sum_{I=1}^3 X^I |_\pm\,.
\end{align}
We can now use this to obtain expressions for the weights $\epsilon_p$ at the fixed points after recalling that, by definition, 
$\dd \xi^\flat |_p=2\epsilon_p\vol$ as well as using \eqref{kvcondLS} to find 
\begin{align}\label{RsymcsLS}
\sum_{I=1}^3\newc^I_+=2-\frac{b_0}{n_+}P_+\,,\qquad
\sum_{I=1}^3\newc^I_-=2+\frac{b_0}{n_-}P_-\,,
\end{align}
which, combined with \eqref{LSBVFLUX}, gives 
\begin{align}\label{RsymconstLS}
\sum_{I=1}^3 \Pnew^I=\frac{n_-P_++ n_+ P_-}{n_- n_+}\,.
\end{align}
all as in the STU model without the hypermultiplet.

The localization of the $D=2$ action proceeds exactly as in the STU model and we again find
\begin{equation}\label{LSSPS}
\begin{aligned}
    S_2|_{\text{POS}} &=\frac{4\pi}{b_0} \left[\mathcal{F}(\newc^I_+)-\mathcal{F}(\newc^I_-)\right]\,.
\end{aligned}
\end{equation}
The action depends on seven variables, $b_0,\newc^I_\pm$. We can eliminate
$\newc^I_-$ using \eqref{LSBVFLUX}. Also, for given 
spindle data $n_\pm$ and $P_\pm$, there is only one independent flux due to 
\eqref{brokenLSflux} and \eqref{RsymconstLS}, which we can take to be $\Pnew_F\equiv \Pnew^1-\Pnew^2$. We also write
$\Pnew_F\equiv p_F/(n_+n_-)$.
Similarly, there is only one independent $\newc^I_+$ due to
\eqref{brokencsLS} and \eqref{RsymcsLS}, which we can take to be $\newc_+^1$.

Thus, for given spindle data $n_\pm$, choice of spinor chiralities 
at the poles $P_\pm$, and independent flux parameter $\Pnew_F$, we need
to vary the action over two variables, which we can take to be $\newc_+^1$ and $b_0$. Extremizing then
leads to the central charge, as well as the value of the warp factor and scalars in the vector multiplet at the poles. Notice, 
however, that this procedure does not fix the value of the complex scalar $\vpvar$ in the hypermultiplet at the poles of the spindle. This exactly mirrors the results found in \cite{Arav:2022lzo} which were obtained by a direct analysis of the BPS equations.

In order to compare to the results of \cite{Arav:2022lzo}  we now take 
the anti-twist case 
$P_+=1,P_-=\sigma=-1$, and $n_\pm\ge 1$, and extremize over $\newc^1_+, b_0$. The extremal value of $b_0$ is given by
\begin{align}
b_0=\frac{n_+n_-(n_++n_-)\left[3(n_+-n_-)^2+4p_F^2\right]}{(n_+-n_-)^2(n_+^2+n_+n_-+n_-^2)+4n_+n_-p_F^2}\,,
\end{align}
and we can also explicitly write down $\newc^1_+$ and hence all of the $\newc^I_\pm$. The extremal value for the action allows us to
write down the on-shell central charge from \eqref{cLScaseitofa}
\begin{equation}
    c=\frac{3(n_--n_+)[(n_+-n_-)^2-4p_F^2][3(n_+-n_-)^2+4p_F^2]}{32n_+n_-[(n_+-n_-)^2(n_+^2+n_+n_-+n_-^2)+4n_+n_-p_F^2]}N^2\,.
\end{equation}
Notice from \eqref{cIdefLS} that with $P_+=1,P_-=\sigma=-1$ we have $\newc^I_-<0$,  $\newc^I_->0$ and we also have $c>0$. From the explicit expressions we find that these conditions are satisfied if and only if 
\begin{align}
n_--n_+>2|p_F|\,,
\end{align}
and $b_0>0$. The flux conditions require that $n_--n_+$ is even and that $p_3,p_F$ are both even or both odd.

For the case of the twist, with $P_+/P_-=\sigma=+1$ but with 
potentially $P_+=P_-=1$ or $P_+=P_-=-1$, 
we can similarly extremize over $b_0,\newc^1_+$ and obtain
the extremal values for $b_0$,  $\newc^I_\pm$ and $c$. Imposing
$\{\newc^I_+<0, \newc^I_-<0,c>0\}$ or $\{\newc^I_+>0, \newc^I_->0,c>0\}$
leads to no solutions.

All of these results are in precise alignment with those\footnote{Note the typo in (3.56) of
\cite{Arav:2022lzo}, \emph{viz.} there should be no factor of $n_N n_S$ on the right hand side.}
found in \cite{Arav:2022lzo}, that were also obtained without 
explicitly solving the BPS equations, but instead just analysing their structure. The results here thus provide an elegant
rederivation using the calculus of equivariant localization.

\section{\texorpdfstring{$D=5$, $\mathcal{N}=2$}{D=5 N=2} ungauged supergravity}\label{sec:d5ungauged}

Starting from the general analysis in section \ref{sec:d5sugramodelvectors}, 
it is straightforward to  obtain results for ungauged 
$D=5$, $\mathcal{N}=2$ supergravity coupled to arbitrary 
vector multiplets,  by simply turning 
off the FI gauging parameters $\xi_I=0$.  
Although we shall only recover known results, we do so 
in a novel way that is likely to generalize further.\footnote{For the specific model with
hypermultipets considered in section \ref{sec:d5sugramodelhypers}, we find that switching off the gauging in
this sector by taking $\zeta_I=0$ implies $ \dd_\xi D\theta=0$ from \eqref{LSdthetacon} and hence one no longer
deduces the constraints \eqref{brokenLSflux} and \eqref{brokencsLS}.
}

We begin by noting that setting $\xi_I=0$ means that the superpotential
$W\equiv 0$ in \eqref{W5d}, and correspondingly also the potential 
$\mathcal{V}\equiv 0$ in \eqref{P5d}. Making the same 
$AdS_3\times M_2$ ansatz \eqref{ansatz5d}, everything goes through 
as written and we may localize on $M_2\cong \mathbb{WCP}^1_{[n_+,n_-]}$. The 
R-symmetry flux in \eqref{twistme} is however now zero, which gives the equation
\begin{align}
n_- P_+ + n_+ P_- = 0\, .
\end{align}
Since $n_\pm\geq 1$ and $|P_\pm|=1$ we necessarily have (without loss 
of generality) $P_+=1$, $P_-=-1$ and $n_+=n_-=n$. 
But then $M_2=S^2/\Z_n$ and we may always lift to the simply-connected 
covering space  $S^2$ with $n=1$, which we henceforth do.  In particular we immediately 
deduce that supersymmetric spindle solutions do not exist in this 
ungauged supergravity theory, which is a new result.

Next the constraints \eqref{cIpm} immediately force $b_0=2$, 
and thus the R-symmetry Killing vector $\xi=2\partial_{\varphi}$ is fixed to rotate 
the $M_2=S^2$ horizon with weight 2. 
The partially off-shell action now reads
\begin{align}\label{SPOSungauged}
    S_2|_{\text{POS}} &=2\pi \left[\mathcal{F}(\newc^I_+)-\mathcal{F}(\newc^I_-)\right]\, ,
\end{align}
where
we have defined
\begin{align}
\newc^I_+ = \newc^I - \Pnew^I\, , \qquad \newc^I_- = \newc^I + \Pnew^I\, ,
\end{align}
and $S_2|_{\text{POS}}$ is to be extremized over the constant variables $\newc^I$. 
Using the form of the prepotential \eqref{constraint}, the extremal equations read
\begin{align}\label{Cxp}
C_{IJK}x^J \Pnew^K = 0\, ,
\end{align}
for each $I$, where recall that $C_{IJK}$ is totally symmetric. 
Substituting \eqref{Cxp} back into 
\eqref{SPOSungauged} we may immediately compute the 
on-shell central charge
\begin{align}\label{cFinn}
c = -\frac{3\pi}{2G_5}\frac{1}{b_0}\left[\mathcal{F}(\newc^I_+) - 
\mathcal{F}(\newc^I_-)\right] = \frac{\pi}{4G_5}C_{IJK}\Pnew^I \Pnew^J\Pnew^K\,.
\end{align}

We may read the equation \eqref{Cxp} as 
 saying that $x^J$ is in the null-space of the 
symmetric matrix $C_{IJK}\Pnew^K$. If the latter is invertible then certainly
the only solution is $x^I=0$. On the other hand, 
notice that 
\eqref{eq:IntroductioncI} gives the values of the physical scalars 
$X^I$ at the fixed poles of the $S^2$ as
\begin{align}
X^I|_{\pm}\, = \mp (\newc^I\mp \Pnew^I )\ex^{-\lambda}|_{\pm}\, .
\end{align}
For solutions where the warp factor and scalars 
are independent of the $S^2$ coordinates, then we immediately see that 
this assumption forces the $x^I=0$ solution above.
Proceeding with this solution, 
the constraint \eqref{constraint} then fixes
\begin{align}\label{elambda}
\ex^{3\lambda}|_{\pm}\,  = \frac{1}{3!}C_{IJK}\Pnew^I \Pnew^J \Pnew^K\, ,
\end{align}
and we deduce
\begin{align}\label{XIpm}
X^I|_{\pm} \, = \frac{\Pnew^I}{(\tfrac{1}{3!}C_{IJK}\Pnew^I \Pnew^J \Pnew^K)^{1/3}}\, ,
\end{align}
which is a well-known formula that may be obtained from the standard
attractor mechanism in $D=5$ for supersymmetric black strings when we have a round metric on $S^2$; 
see, for example, \cite{Larsen:2006xm}. 

This theory is the low-energy effective theory one obtains by 
compactifying M-theory on a Calabi--Yau three-fold $X_3$. 
Including the graviphoton, the number of Abelian vector multiplets 
is $h^{1,1}$, while the number of hypermultiplets (here turned off)
is $h^{2,1}+1$, where $h^{i,j}=h^{i,j}(X_3)$ denote Hodge numbers. 
Introducing a basis of two-cycles $\Omega^I$, $I=1,\ldots,h^{1,1}$, generating the 
free part of $H_2(X_3,\Z)$, and a dual basis of four-cycles 
$\Omega_I$ with intersection numbers $\Omega^I\cap \Omega_J =\delta^I_J$, 
then the scalar fields are
\begin{align}
X^I = \int_{\Omega^I} \mathcal{J}\, , 
\end{align}
where $\mathcal{J}$ is the K\"ahler form on the Calabi--Yau three-fold $X_3$. 
The $C_{IJK}$ are the intersection numbers
\begin{align}
C_{IJK} = \Omega_I\cap \Omega_J\cap\Omega_K\, .
\end{align}
The supersymmetric extremal black string solutions 
we have discussed above arise by wrapping an M5-brane over the four-cycle
$\Pnew^I\Omega_I$, and we see that the central charge 
\eqref{cFinn} simply counts the triple intersection number of 
this wrapped M5-brane. 

More generally in our approach nothing forces the warp factor 
and scalars to be constant over the horizon, and 
depending on $C_{IJK}$ one may wonder if there are more general $AdS_3\times S^2$ solutions
than the homogeneous choice $x^I=0$ above; in particular, is it possible\footnote{We thank James Lucietti for discussion on this.}
 to have solutions with an inhomogeneous metric on the $S^2$? 
We have assumed that there is at least an axial symmetry on the 
$S^2$, so in the case of minimal $\mc{N}=2$ ungauged supergravity it is known \cite{Kunduri:2009ud} that the only possibility is
$AdS_3\times S^2$ with a round $S^2$; it would be interesting to extend \cite{Kunduri:2009ud} to
include vector multiplets. We can also consider the uniqueness results of \cite{Reall:2002bh,Gutowski:2004bj};
since any $AdS_3\times M_2$ can be interpreted as a near horizon geometry with horizon $S^1\times M_2$, by writing
$AdS_3$ in Gaussian null coordinates, the results of \cite{Reall:2002bh,Gutowski:2004bj} imply that the metric on $S^2$ should be the round metric, provided that the R-symmetry Killing vector is null on a hypersurface on  $AdS_3$. This leaves the possibility of 
solutions with non-round metrics on $S^2$ if the R-symmetry Killing vector is strictly timelike, which is
not something we have considered in our analysis. It would be interesting if this argument could be extended to conclude that any 
supersymmetric $AdS_3\times S^2$ solution must have a round metric on the $S^2$.

\section{\texorpdfstring{$D=4$, $\mathcal{N}=2$}{D=4 N=2} gauged supergravity}
\label{sec:D4SUGRA}

In this section we discuss  $\mc{N}=2$ gauged supergravity in four dimensions, coupled to Abelian vector multiplets \cite{deWit:1984wbb}. Note that we slightly modify the normalization of the gauge fields and gauge coupling compared to \cite{Freedman:2012zz, Bobev:2020pjk} (see \cite{Cacciatori:2008ek} as well for a review). 

The bosonic content of the graviton supermultiplet is the metric and a graviphoton $A^0$, while that of the $n$ Abelian vector multiplets
consists of gauge fields $A^i$ and complex scalar fields $z^i$, where $i=1, \dots, n$. It is convenient to introduce the index $I=0,1, \dots, n$ to unify the description of the gauge fields. The scalars parametrize a special K\"ahler manifold of complex dimension $n$, with  K\"ahler potential $\cK$ and metric $\mc{G}_{i\overline{j}} = \partial_i \partial_{\overline{j}}\cK$, that is the base of a symplectic bundle with covariantly holomorphic sections
\begin{equation}
	\ex^{\cK/2} \begin{pmatrix}
		X^I \\ \mc{F}_I
	\end{pmatrix} \, .
\end{equation}
Here the scalars $X^I=X^I(z^i)$. 
These obey a symplectic constraint that allows us to express the K\"ahler potential in terms of the sections
\begin{equation}
\label{eq:4d_SymplecticConstraint}
	\ex^{\cK} \left( X^I \overline{\mc{F}}_I - \mc{F}_I \overline{X}^I \right) = \ii \qquad \Rightarrow \qquad \cK = - \log \left[ \ii \left( \mc{F}_I \overline{X}^I - X^I \overline{\mc{F}}_I \right) \right] \, .
\end{equation}
We shall assume there exists a holomorphic prepotential $\mc{F}=\mc{F}(X^I)$, which is homogeneous of degree two,
such that $\mc{F}_I = \frac{\partial \mc{F}}{\partial X^I}$.\footnote{In general the existence of the prepotential depends on the choice of symplectic sections, see \cite{Ceresole:1995jg}. 
} We define
\begin{equation}
	\mathcal{N}_{IJ} = \overline{\mc{F}}_{IJ} + \ii \frac{N_{IK} X^K N_{JL} X^L}{N_{NM} X^N X^M} \, ,
\end{equation}
where $\mc{F}_{IJ} = \partial_I \partial_J \mc{F}$ and $N_{IJ} = 2 \Im \mc{F}_{IJ}$. 

The bosonic part of the action, which is determined by the prepotential, is given by
\begin{align}
\label{eq:4d_Action}
	S_4 = \frac{1}{16\pi G_4} \int_{M_4} \bigg( & R_4 + \frac{1}{4} \Im \mc{N}_{IJ} F^I_{\mu\nu} F^{J\mu\nu} + \frac{1}{4\ii} \Re \mc{N}_{IJ} F^I_{\mu\nu} \widetilde{F}^{J\mu\nu} \nn \\
	& - 2\mc{G}_{i \overline{j}} \partial^\mu z^i \partial_\mu \overline{z}^{\overline{j}} - \mc{V} \bigg) \vol_4 \, ,
\end{align}
where $\widetilde{F}^{I\mu\nu} \equiv - \frac{\ii}{2} \epsilon^{\mu\nu\rho\sigma} F^I_{\rho\sigma} $.
Here we have assumed that there is no gauging of the special K\"ahler isometries, and the potential is given by
\begin{align}
\label{eq:4d_ScalarPotential}
	\mc{V} \equiv \xi_I \xi_J \ex^{\cK} \left(\mc{G}^{i \overline{j}} \nabla_i X^I \nabla_{\overline{j}}\overline{X}^J - 3 X^I \overline{X}^J\right) \, ,
\end{align}
where the K\"ahler covariant derivatives of $X^I$ are
\begin{equation}
	\nabla_i X^I \equiv \left( \partial_i + \partial_i \cK \right) X^I \, ,
\end{equation}
and we have chosen to work with $U(1)$ Fayet--Ilioupoulos gauging with $\xi_I\in \R$. It is useful to introduce the holomorphic superpotential
\begin{equation}
	W \equiv \xi_I X^I \, ,
\end{equation}
in terms of which the scalar potential is written as
\begin{equation}
\label{eq:4d_ScalarPotential_v2}
	\mc{V} = \ex^\cK \left( \mc{G}^{i \bar{j}} \nabla_i W \nabla_{\bar{j}}\overline{W} - 3 W \overline{W} \right) \, .
\end{equation}
For use in later sections, it is also helpful to define the real superpotential
\begin{equation}
\label{eq:4d_RealSuperpotential}
	\mc{W} = - \sqrt{2} \ex^{\mc{K}/2} \abs{W} \, ,
\end{equation}
and the potential then reads
\begin{equation}
\label{eq:4d_ScalarPotential_v3}
	\mc{V} = 2 \mc{G}^{i \bar{j}} \partial_i \mc{W} \partial_{\bar j} \mc{W} - \frac{3}{2} \mc{W}^2 \, .
\end{equation}

A solution is supersymmetric if there exists a non-vanishing Dirac spinor $\epsilon$ satisfying the following equations\footnote{The supersymmetry variations of the gravitino and the gauginos are commonly written in terms of two Weyl spinors of definite chirality. For our purposes it is more convenient to combine them in a single Dirac spinor, as done in \cite{Cacciatori:2008ek} 
(and recall we have modified the normalization of the gauge fields and gauge coupling).}
\begin{align}
	0 &= \nabla_\mu \epsilon +  \frac{\ii}{2} \mc{A}_\mu  \Gamma_5 \epsilon - \frac{\ii}{4} \xi_I A_\mu^I \epsilon + \frac{1}{2\sqrt{2}} \Gamma_\mu \ex^{\mc{K}/2} \left( \Im W + \ii \Re W \, \Gamma_5 \right) \epsilon \nn \\
\label{eq:4d_GravitinoVariation}
	& \ \ \ - \frac{\ii}{4\sqrt{2}} \Im \mc{N}_{IJ} F^{J}_{\nu\rho} \Gamma^{\nu\rho} \Gamma_\mu \ex^{\mc{K}/2} \left( \Im X^I + \ii \Re X^I \, \Gamma_5 \right) \epsilon \, , \\[5pt]
	0 &= \frac{1}{2\sqrt{2}} \Im \mc{N}_{IJ} F^{J}_{\mu\nu} \Gamma^{\mu\nu} \ex^{\mc{K}/2} \left[ \ii \Im ( \mc{G}^{\bar{i} {j}} \nabla_j X^I) - \Re ( \mc{G}^{\bar{i} {j}}  \nabla_j X^I) \, \Gamma_5 \right] \epsilon \nn \\
\label{eq:4d_GauginoVariation}
	& \ \ \ + \Gamma^\mu \left( \Re \partial_\mu z^i - \ii \Im \partial_\mu z^i \, \Gamma_5 \right) \epsilon \nn \\ 
& \ \ \ + \frac{\ii}{\sqrt{2}} \ex^{\mc{K}/2} \left[\ii \Im ( \mc{G}^{\bar{i} {j}}  \nabla_j W) - \Re ( \mc{G}^{\bar{i} {j}}  \nabla_j W) \, \Gamma_5 \right] \epsilon \, ,
\end{align}
where $\Gamma_\mu$ generate Cliff$(1,3)$ and $\Gamma_5 \equiv \ii \Gamma_{0123}$, and the $U(1)$ connection $\mc{A}$ is defined by
\begin{equation}
\label{eq:KahlerHodgeConnection}
	\mathcal{A}_\mu = - \frac{\ii}{2}\left(\partial_i \mathcal{K} \partial_\mu z^i - \partial_{\bar{i}} \mathcal{K} \partial_\mu \overline{z}^{\bar{i}}\right) \, .
\end{equation}

\subsection{The \texorpdfstring{$AdS_2$}{AdS2} ansatz}
\label{subsec:AdS2Ansatz}

We consider supersymmetric $AdS_2\times M_2$ solutions with a warped product metric
\begin{equation}\label{eq:4d_Ansatz}
     \dd s_4^2 = \ex^{2\lambda}\left[\dd s^2(AdS_2)+\dd s^2(M_2)\right]\,. 
\end{equation}
The metric on $AdS_2$ has unit radius, $\lambda$ and the scalar fields are functions on $M_2$, and the gauge fields $A^I$ are (pull-backs of) gauge fields on $M_2$. These solutions can arise as the near horizon limit of magnetically charged static extremal supersymmetric black holes.

We can now proceed as in section \ref{sec:AdS3}. We first insert the ansatz in the $D=4$ equations of motion, obtaining $D=2$ equations of motion, which can be obtained from the effective action (which can also be obtained by substituting the ansatz directly in \eqref{eq:4d_Action})
\begin{align}
	S_2 = \int_{M_2} \bigg[ &\ex^{2\lambda} \left( R - 2 + 6 (\nabla\lambda)^2 - 2\mc{G}_{i \overline{j}} \partial^\mu z^i \partial_\mu \overline{z}^{\overline{j}} \right) - \ex^{4\lambda} \mc{V} \nn \\
\label{eq:4d_2dEffective}
	& + \frac{1}{4} \Im \mc{N}_{IJ} F^I_{\mu\nu}F^{J\mu\nu} \bigg] \vol \, ,
\end{align}	
where $R$ is the Ricci scalar of $M_2$, and $\nabla$ and $\vol$ are the Levi--Civita connection and the volume form. We are interested in the Maxwell equation for $A^I$ and the equation for $\lambda$, which read
\begin{align}
	\label{eq:4d_Maxwell}
	\dd \left( \Im \mc{N}_{IJ} * F^J \right) = 0 \quad \Rightarrow \quad \dd \left( \Im \mc{N}_{IJ} F^J_{12} \right) = 0 \, , \ \forall I \, ,  \\
	\label{eq:4d_lambda}
	\ex^{2\lambda} \left( R - 2 + 6 (\nabla\lambda)^2 - 2\mc{G}_{i \overline{j}} \partial^\mu z^i \partial_\mu \overline{z}^{\overline{j}} \right) = 2 \ex^{4\lambda} \mc{V} + \text{total derivative}.
\end{align}
Imposing the equation for the warp factor leads to the following partially off-shell two-dimensional action
\begin{equation}
\label{eq:4d_ReducedS}
	 S_2|_{\text{POS}} = \int_{M_2} \left( \ex^{4\lambda} \mc{V} \, \vol + \frac{1}{2} \Im \mc{N}_{IJ} F^I_{12} F^J \right) \, .
\end{equation}
The trace of the Einstein equation derived from the action $S_2$ is
\begin{equation}
	\ex^{4\lambda} \mc{V} + \frac{1}{4} \Im \mc{N}_{IJ} F^I_{\mu\nu} F^{J\mu\nu} = - 2\ex^{2\lambda} \, ,
\end{equation}
and imposing this gives us an expression for the on-shell action
\begin{equation}
	S_{2} \rvert_{\rm OS} = - 2\int_{M_2} \ex^{2\lambda} \, \vol \, .
\end{equation}
If  \eqref{eq:4d_Ansatz} is the near horizon limit of an extremal black hole, then $S_2\rvert_{\rm OS}$ is related to its Bekenstein--Hawking entropy
\begin{equation}
\label{eq:4d_SBH}
	S_{\rm BH} = - \frac{1}{8G_4} S_2 \rvert_{\rm OS} \, .
\end{equation}
We will compute this quantity for arbitrary $M_2$ without explicitly solving the equations of motion, but instead defining an off-shell entropy function
\begin{equation}
\label{eq:4d_SBH_OffShell}
	S_{\rm BH} = - \frac{1}{8 G_4} S_2 \rvert_{\text{POS}} \, ,
\end{equation}
which we compute using the BVAB localization theorem. We can then extremize over the remaining parameters in order to get $S_{2} \rvert_{\rm OS}$, and thus the entropy of the black hole.

\medskip

What we just showed suggests a relation with Euclidean quantum gravity. Recall that $S_2\rvert_{\text{POS}}$ can be obtained by imposing the trace of the Einstein equations of motion in $S_4$ \eqref{eq:4d_Action}, so one has the following
\begin{equation}
\label{eq:4d_Thermodynamics}
	S_{\rm BH} = - \frac{2\pi}{\Vol(AdS_2)} S_4\rvert_{\text{trace $E_{g_4}$}} = \frac{2\pi}{\Vol(\mathbb{H}_2)} S_4^E\rvert_{\text{trace $E_{g_4}$}} = - S_4^E \rvert_{\text{trace $E_{g_4}$}} \, .
\end{equation}
Here, the first equality follows from the definition, the second one is just a Wick rotation, and the third one uses the renormalized volume of Euclidean $AdS_2$. Therefore, at least at the formal level, we find a relation between the entropy function of the black hole and the four-dimensional Euclidean on-shell action. Of course, this relation is suggestive of the canonical relation between the entropy and the Euclidean on-shell action for magnetically charged black holes: $S_{\rm BH} = - I$. However, there are various caveats. First, in order to establish the latter we would have to include the Gibbons--Hawking--York term and the holographic renormalization counterterms, which are not known for an arbitrary gauged supergravity model. Moreover, notice that \eqref{eq:4d_Thermodynamics} is a relation between the entropy and the action of the \emph{near horizon} geometry $AdS_2\times M_2$, whereas the thermodynamics would require the on-shell action of the entire black hole spacetime. This concern is in fact addressed by the equivariant localization itself. As we shall see, $S_{\rm BH}$ is computed using the BVAB theorem, and therefore receives contributions only from fixed point sets, with the result depending only on the topology of spacetime. It is then conceivable that the extremal black hole geometry arises from a limit of a family of geometries with topology disc$\times M_2$, in which case the action of the entire family of solutions would be the same, independent of the limit parameter. Whilst this is only a conjecture for the model currently studied, it can be proved for the case of minimal
$\mathcal{N}=2$ gauged supergravity, both with spindle \cite{Cassani:2021dwa} and with Riemann surface horizon \cite{BenettiGenolini:2019jdz}.\footnote{In the latter case, namely universal black holes with Riemann surface horizon, the relation $S_{\rm BH} = - I$ has also been established by finding families of solutions both with supersymmetry \cite{BenettiGenolini:2023ucp} and which break supersymmetry \cite{Azzurli:2017kxo, Halmagyi:2017hmw, Cabo-Bizet:2017xdr}. The existence of supersymmetric deformations of the black hole with closed interior has been discussed within the STU model in \cite{Bobev:2020pjk}.} At the same time, this also resolves the concern over the divergence of the on-shell action due to the $AdS_2$ infinite throat. We leave solving these issues and establishing the suggestive relation \eqref{eq:4d_Thermodynamics} to future work.

\subsection{Equivariantly closed forms and localization}
\label{subsec:4d_Generic_Localization}

Following the same procedure as in section \ref{sec:5dequivariant}, we impose a spinor ansatz consistent with \eqref{eq:4d_Ansatz}, namely
\begin{equation}
\label{eq:4d_SpinorAnsatz}
	\epsilon = \vartheta \otimes \ex^{\lambda/2}\zeta \, , 
\end{equation}
where $\zeta$ is a spinor on $M_2$, and now $\vartheta$ is a Killing spinor on $AdS_2$. This allows us to reduce the spinor equations \eqref{eq:4d_GravitinoVariation} and \eqref{eq:4d_GauginoVariation} to spinor equations on $M_2$, as we show in appendix \ref{app:4dBilinears}. We only have magnetic fields on $M_2$, so in particular $F^I\wedge F^J=0$; it is then consistent to
now set the symplectic sections $X^I$ to be purely imaginary\footnote{\label{footu1}Using a $U(1)$ rotation, 
associated with the gauge field \eqref{eq:KahlerHodgeConnection}, it is possible
to change the phase of the symplectic section if we correspondingly rotate the Killing spinors, 
as discussed around (4.16) in \cite{Hristov:2010ri} (see also \cite{Cabo-Bizet:2017xdr}).
We also highlight that in section \ref{sec:4dSTU} we will in fact utilize such a rotation to discuss the STU model. More details are in appendix \ref{app:4dBilinears}. \label{footnote:Rsymmetry}} 
\begin{equation}
\label{eq:4d_Constraint}
	z^i \in \R \, , \qquad X^I \in \ii \R \, .
\end{equation}

As shown in appendix \ref{app:4dBilinears}, we then define the same real bilinears as in \eqref{5dbilinears}, and again a canonical analysis of the bilinear equations leads to the conclusion that $\xi$ is a Killing vector, and that the following polyforms are equivariantly closed
\begin{align}
\label{eq:4d_EquivariantPolyforms}
	\Phi^{F^I} &\equiv F^I + 2 \sqrt{2} \ex^\lambda \ex^{\cK/2} \ii X^I P \, , \nn\\
	\Phi^{\rm vol} &\equiv \ex^{4\lambda} \, \mc{V} \, \vol + \sqrt{2}\ex^{3\lambda} \ex^{\mc{K}/2} \ii W S \, ,
\end{align}
where the bilinear $S$ is a constant that we henceforth set to $1$.
We can then immediately define another equivariantly closed form $\Phi^S$ given by
\begin{align}\label{phisdefn4}
\Phi^S=\Phi^{\rm vol} + \frac{1}{2} \Im \mc{N}_{IJ}F^I_{12} \Phi^{F^J}\,.
\end{align}

Having constructed the equivariantly closed polyforms, and assuming that $\xi$ acts on $M_2$ with only isolated fixed points, then we may appeal to the Berline--Vergne--Atiyah--Bott theorem to evaluate their integrals, as done in section \ref{sec:5dLocalization}. Indeed, we find the magnetic charges\footnote{In general there is no reason to quantize the magnetic charges. For models that can be uplifted to string/M-theory, such as the STU model discussed
in section \ref{sec:4dSTU}, the suitably normalized gauge fields are associated with $U(1)$ bundles and there is a quantization condition. \label{footnote:quantization}}
\begin{align}
\label{eq:4d_MagneticCharges}
	\Pnew^I &= \int_{M_2} \frac{\Phi^{F^I}}{4\pi} = \frac{1}{4\pi} \sum_{\mathrm{fixed\, } p} \frac{1}{d_p} \frac{2\pi}{\epsilon_p} ( 2\sqrt{2} \ex^\lambda \ex^{\cK/2} \ii X^I P ) \rvert_p \, ,
\end{align}
and action
\begin{align}\label{eq:4d_SBV_v1}
	S_2 \rvert_{\text{POS}} &= \int_{M_2}\Phi^S\nn\\
	&= \sum_{\mathrm{fixed\, } p} \frac{1}{d_p} \frac{4\pi}{\epsilon_p} \left. \frac{1}{\sqrt{2}} \left( \ex^{3\lambda} \ex^{\mc{K}/2} \ii W S + \ex^\lambda \Im \mc{N}_{IJ} F^I_{12} \ex^{\mc{K}/2} \ii X^J P \right) \right\vert_p\nn \\
	&= - \sum_{\mathrm{fixed\, } p} \frac{1}{d_p} \frac{4\pi}{\epsilon_p} P\,  \ex^{2\lambda} \rvert_p \, .
\end{align}	
Analogously to the evaluation of \eqref{actionp}, we used the Maxwell equation \eqref{eq:4d_Maxwell} and the algebraic relation \eqref{eq:4d_AlgebraicF12} obtained from the spinor equations.

Finally, we introduce the scalars $\newc^I$ that are the degree-zero components of the equivariant Chern classes \eqref{eq:4d_EquivariantPolyforms}, namely
\begin{equation}
	\newc^I \equiv - 2\sqrt{2} \ex^\lambda \ex^{\mc{K}/2} \ii X^I P \, .
\end{equation}
In terms of these, the equation \eqref{eq:4d_dxib} for $\dd\xi^\flat$ reads
\begin{equation}
\label{eq:4d_Constraint_BV}
	\dd\xi^\flat = 2 \left( 1 - \frac{1}{2} \xi_I \newc^I \right) P \, \vol \quad \Rightarrow \quad \xi_I \newc^I_p = 2 - 2 P_p \epsilon_p \, ,
\end{equation}
and the magnetic charges are
\begin{equation}
\label{eq:4d_PI_v1}
	\Pnew^I = - \frac{1}{2} \sum_{\mathrm{fixed\, } p} \frac{1}{d_p \epsilon_p} \newc^I_p \, .
\end{equation}
To conclude, we look again at the symplectic constraint \eqref{eq:4d_SymplecticConstraint} under the reality assumptions \eqref{eq:4d_Constraint}. This is only consistent if $\mc{F}(X^I)$ is pure imaginary, and using homogeneity of $\mc{F}$ leads to
\begin{equation}
\label{eq:4d_eKFX_Constraint}
	\ex^{\mc{K}} \ii \mc{F}(X^I) =  -\frac{1}{4} \, .
\end{equation}
Using again homogeneity, we find that
\begin{equation}
\label{eq:4d_eKFX_Constraint_2}
	\ii \mc{F}(\newc^I) \rvert_p = 2 \ex^{2\lambda} \rvert_p \, ,
\end{equation}
and thus the reduced action \eqref{eq:4d_SBV_v1} takes the form
\begin{equation}
\label{eq:4d_SBV_v2}
	S_2 \rvert_{\text{POS}} = - \sum_{\mathrm{fixed\, } p} \frac{2\pi}{d_p \epsilon_p} P_p\,  \ii \mc{F}(\newc_p^I) \, .
\end{equation}
Notice that \eqref{eq:4d_PI_v1} and \eqref{eq:4d_SBV_v2} have the same form as their analogues \eqref{cIpformulas}. Therefore, for the four-dimensional solutions of type \eqref{eq:4d_Ansatz}, again we can compute the action and the magnetic charges knowing only the values of the scalars $\newc^I_p$ at the fixed points, together with the data of the circle action, namely $d_p$, $\epsilon_p$ and $P_p$.

Having established these results, we now consider $M_2 \cong \mathbb{WP}_{[n_+, n_-]}$ to be a spindle.
There are two fixed points labelled by $\pm$ with the order of the orbifold group $d_\pm = n_\pm$, where we take $n_\pm\ge 1$,
and weights $\epsilon_\pm = \mp b_0 / n_\pm$, where $b_0$ is the weight of the azimuthal circle action. We also set  
$P_- = \sigma P_+$. 
After application of the BV theorem the
form of the integrals for $\Pnew^I$ and $S_2\rvert_{\text{$\lambda$ EOM}}$ 
are the same as in section \ref{sec:5dLocalization}, so we can now follow the same steps as we did there.

We find that the magnetic charges \eqref{eq:4d_PI_v1} are given by
\begin{equation}
\label{eq:4d_PI_v2}
	\Pnew^I =  \frac{1}{2b_0} ( \newc^I_+ - \newc^I_-) \, ,
\end{equation}
and the constraint \eqref{eq:4d_Constraint_BV}, which now reads
\begin{equation}
\label{eq:4d_xIConstraintsGeneral}
	\xi_I x^I_+ = 2 + P_+ \frac{2b_0}{n_+} \, , \qquad \xi_I x^I_- = 2 - \sigma P_+ \frac{2b_0}{n_-}\,,
\end{equation}  
leads us to an expression for the total magnetic flux of the R-symmetry through the spindle of the form
\begin{equation}
\label{eq:4d_RSymmetryFlux}
	\xi_I \Pnew^I = P_+ \frac{n_- + \sigma n_+}{n_- n_+} \, .
\end{equation}
Similarly, the off-shell entropy function obtained from the reduced action takes the form
\begin{equation}
\label{eq:filippo}
	S_{\rm BH} = - \frac{1}{8G_4} S_2 \rvert_{\text{POS}} = - \frac{\pi}{2 G_4} \frac{1}{2 b_0} P_+ \left[ \ii\mc{F}(\newc_+^I) - \sigma \ii \mc{F}(\newc_-^I) \right] \, .
\end{equation}
As in the discussion of \eqref{actioncIpm}, these constraints imply that we have a variational problem with $n$ degrees of freedom. 

We can also consider the limit that $b_0\to 0$ in order to obtain non-trivial results for the case $M_2=S^2$.
Specifically, we can substitute $\newc^I_-$ using \eqref{eq:4d_PI_v2} and take the $b_0\to 0$ limit in the $\sigma = 1$ case to get
\begin{equation}
\label{eq:4d_SBH_S2Limit}
	\lim_{b_0 \to 0} S_{\rm BH} = - P_+ \frac{\pi}{2 G_4} \Pnew^I \frac{\partial \ }{\partial \newc^I} \ii\mc{F}(\newc^I) \, ,
\end{equation}
where note that $\newc_+^I = \newc_-^I \equiv \newc^I$ and they are subject to the constraint $\xi_I\newc^I=2$. In the case $n_+ = n_- = 1$, the constraint \eqref{eq:4d_RSymmetryFlux} 
on the fluxes reads $\xi_I \Pnew^I = 2 P_+$ and is associated with the standard topological twist on $S^2$.
This expression for the off-shell entropy has also been obtained by expressing the BPS equations of gauged supergravity as attractor equations
\cite{Cacciatori:2009iz,DallAgata:2010ejj}. 

\subsection{STU model}\label{sec:4dSTU}
The four-dimensional gauged STU model is an $\mc{N}=2$ supergravity theory containing $n=3$ vector multiplets with $\xi_I = 1$ for all values of $I$, and prepotential
\begin{equation}
\label{eq:STUPrepotential}
	\ii\mc{F} = 2\sqrt{X^0 X^1 X^2 X^3} \, .
\end{equation}
It is a consistent truncation of eleven-dimensional supergravity on $S^7$, where the Cartan subgroup $U(1)^4 \subset SO(8)$ is gauged \cite{Azizi:2016noi}.

In contrast to the choice made previously in \eqref{eq:4d_Constraint}, for the STU model we now choose
\begin{equation}
\label{eq:4d_RealityConstraint_STU}
	z^i \in \R \, , \qquad X^I \in \R \, ,
\end{equation}
in order to facilitate some comparisons with the literature ({\it e.g.} \cite{Ferrero:2021etw}). Specifically,  as in \cite{Cvetic:1999xp} we choose
\begin{equation}
\label{eq:4d_GaugeFixingSymplecticCondition}
\begin{aligned}
	X^0 &= \ex^{ \frac{1}{2}( \varphi^1 + \varphi^2 + \varphi^3)} \, , &\qquad X^1 &= \ex^{ \frac{1}{2}( \varphi^1 - \varphi^2 - \varphi^3)} \, , \\
	X^2 &= \ex^{ \frac{1}{2}( - \varphi^1 + \varphi^2 - \varphi^3)} \, , &\qquad X^3 &= \ex^{ \frac{1}{2}( - \varphi^1 - \varphi^2 + \varphi^3)} \, .
\end{aligned}
\end{equation}
Here we have used a standard parametrization of the $X^I$ in terms of three
real scalars $\varphi^i$, as opposed to the  scalars $z^i$ used in the general case. 
As already mentioned in footnote \ref{footu1}, and discussed further in appendix \ref{app:4dBilinears},
we can make this alternative choice provided that we carry out a 
$U(1)$ rotation. This rotation introduces some changes to the Killing spinor equation on $M_2$ that one needs to analyse, but otherwise
the analysis of the previous subsection goes through unchanged.
With this new gauge choice, we have $X^0 X^1 X^2 X^3 = 1$, so $\ii\mc{F}= 2$ and $\mc{K} = - \log 8$, so that $\ex^\mc{K} \ii \mc{F}(X^I) = 1 /4$ (in contrast with \eqref{eq:4d_eKFX_Constraint}). 
The conventions we use for the STU model are precisely the same as those used in the next section (upon setting $\rho = \theta = 0$). 
We also recall that the $AdS_4$ vacuum solution of the STU model is dual to the ABJM theory (with $k=1$) and the four-dimensional Newton constant is normalised as
\begin{equation}
\label{eq:4d_G4_Vacuum}
\frac{1}{G_4} = \frac{2\sqrt{2}}{3} N^{3/2} \, ,
\end{equation}
where $N$ is the rank of the gauge group of ABJM.

Following the procedure in the previous sections, specifically \eqref{eq:4d_SBH} and \eqref{eq:filippo}, we find that
the off-shell entropy function for the STU model is given by
\begin{equation}
\label{eq:4d_STU_SBHOffShell}
	S_{\rm BH} = - P_+ \frac{\pi \sqrt{2}}{3} N^{3/2} \, \frac{1}{b_0} \left( \sqrt{\newc_+^0 \newc_+^1 \newc_+^2 \newc_+^3} - \sigma\sqrt{\newc_-^0 \newc_-^1 \newc_-^2 \newc_-^3} \right) \, ,
\end{equation}
together with the constraints \eqref{eq:4d_xIConstraintsGeneral} and \eqref{eq:4d_RSymmetryFlux}, which read
\begin{equation}
\label{eq:4d_Constraints_STU}
	 \sum_{I=0}^3 \newc^I_+ =   2 + P_+ \frac{2 b_0}{n_+}  \, , \qquad \sum_{I=0}^3 \newc^I_- =   2 -  \sigma P_+ \frac{2 b_0}{n_-}  \, , \qquad \sum_{I=0}^3 \Pnew^I = P_+ \frac{n_- + \sigma n_+}{n_- n_+} \, .
\end{equation}
Extremizing this function leads to the entropy of the accelerating black holes, 
with various different subfamilies (and/or near horizon limits) studied in 
 \cite{Ferrero:2020twa, Cassani:2021dwa, Ferrero:2021ovq, Couzens:2021cpk, Boido:2022iye} (see also  \cite{Lu:2014sza} for earlier work on accelerating black holes 
in STU gauged supergravity). The discussion of the most general case is quite involved \cite{Ferrero:2021ovq}, 
although the near horizon solutions have been analysed in detail in \cite{Couzens:2021cpk}. However, it is straightforward to obtain simple analytic formulae in subfamilies of examples,
and here for illustration we look at the simplest case of minimal $\mathcal{N}=2$ gauged supergravity. This is obtained setting $\varphi^i = 0$ for all $i$ and $A^I \equiv A$ for all $I$.  Upon choosing $P_+ = -1$ (we could also choose $P_+=+1$), the off-shell entropy function becomes
\begin{equation}
	S_{\rm BH} = \frac{\pi \sqrt{2}}{3} N^{3/2} \frac{1}{4b_0} \left( 1 - \sigma - 2 b_0 \frac{n_- +  n_+}{n_+ n_-} + b_0^2 \frac{n_-^2 - \sigma n_+^2}{n_+^2 n_-^2} \right) \, ,
\end{equation}
where we have solved the constraints \eqref{eq:4d_Constraints_STU}, and the magnetic charge satisfies
\begin{equation}
	\Pnew = - \frac{n_- + \sigma n_+}{4n_- n_+} \, .
\end{equation}
Extremizing $S_{\rm BH}$ leads to
\begin{equation}
	b_{0\pm} = \pm \frac{n_+ n_- \sqrt{1-\sigma} }{\sqrt{n_-^2 - \sigma n_+^2}} \, .
\end{equation}
Therefore, in the twist case we obtain the extremum at $b_{0*}=0$, a case associated with $S^2$, which we examine further below.
In the anti-twist case we find a positive entropy for $b_{0+}$:
\begin{equation}
	S_{\rm BH} = \frac{\pi \sqrt{2}}{3} N^{3/2} \frac{ \sqrt{2(n_+^2 + n_-^2)} - (n_+ + n_-) }{2 n_+ n_-} \, .
\end{equation}
This reproduces the entropy of the explicit family of supersymmetric, accelerating, magnetically 
charged black holes in \cite{Ferrero:2020twa}. Notice that these anti-twist solutions have
$n_+<b_0<n_-$, so $\Pnew<0$, with
\begin{equation}
	x_\pm = \frac{1}{2} \left( 1 - n_\mp \sqrt{\frac{2}{n_-^2 + n_+^2}} \right) \, , \qquad \ex^{2\lambda} \rvert_\pm = x_\pm^2 \, ,
\end{equation}
and furthermore $x_+<0$, $x_->0$, as required (see \eqref{xprelhyperfour}).

We can also consider taking the $b_0\to 0$ limit, and set $n_\pm = 1 = \sigma$ to recover the $S^2$ case, as done in \eqref{eq:4d_SBH_S2Limit}. Taking this 
limit of \eqref{eq:4d_STU_SBHOffShell}, {with $P_+ = -1$,}
we obtain the 
entropy function
\begin{equation}\label{BZentropy}
	\lim_{b_0 \to 0} S_{\rm BH} = \frac{\pi\sqrt{2}}{3} N^{3/2} \sqrt{\newc^0 \newc^1 \newc^2 \newc^3} \sum_{I=0}^3 \frac{\Pnew^I}{\newc^I} \, .
\end{equation}
The variables are now constrained to obey $\sum_{I=0}^3 \newc^I = 2 = - \sum_{I=0}^3 \Pnew^I$. 
This precisely matches the result of \cite{Benini:2015eyy}. In particular, when comparing with the topologically twisted index 
in the
dual field theory one identifies\footnote{This mapping uses the notation of \cite{Bobev:2018uxk}. To match \cite{Benini:2015eyy}, one needs to rescale $\Delta^I$ by $\pi$ and change the sign of $\mathfrak{n}^I$.}
\begin{equation}\label{identifystuff}
	\Pnew^I \leftrightarrow \mathfrak{n}^I \, , \qquad \newc^I \leftrightarrow \Delta^I \, .
\end{equation}
Here $\mathfrak{n}^I$ are the magnetic fluxes of the background R-symmetry gauge fields, and $\Delta^I$ are chemical potentials for the ABJM vector multiplets. The first identification is clear, being the same flux. The second identification, instead, relies on the existence of the uplift on $S^7$, in which case the values of the $\newc^I$ at the fixed points should be identified with the weights of the R-symmetry vector, and thus with the R-charges, 
as explained in section \ref{sec:UpliftAdS3}. 
Thus, in writing \eqref{identifystuff} we are identifying the chemical potentials $\Delta^I$ for the topologically twisted index 
with R-charges in the gravity dual; this has of course been noticed before,
but the way we have arrived at it here seems to provide a new perspective, 
that might be worth investigating further.

\subsection{Uplifting to \texorpdfstring{$AdS_2\times Y_9$}{AdS2 x Y9} solutions}
\label{sec:UpliftAdS2}

The spindle STU solutions just discussed can be uplifted on $S^7$ to solutions of eleven-dimensional supergravity \cite{Ferrero:2020twa, Ferrero:2021ovq}. The discussion of the uplift and off-shell geometry are quite similar to that in section \ref{sec:UpliftAdS3}, so we shall be brief. 

The eleven-dimensional solution has the form $AdS_2 \times Y_9$, where $Y_9$ is the total space of a $S^7$ fibration over $M_2$. More specifically, in analogy to \eqref{fibredS5}, $S^7$ is the sphere bundle in $\R^8$ acted upon by $U(1)^4$, where each $U(1)$ acts on one of the transverse two-planes in $\R^8$ and the fibration introduces a connection one-form $A^I$ that uplifts the $D=4$ gauged supergravity field. The magnetic charges \eqref{eq:4d_MagneticCharges} are then quantized as $\Pnew^I = p^I/n_+ n_-$, with $p^I \in \Z$, and they correspond to the Chern numbers of the $Y_9$ fibration \cite{Boido:2022mbe}. By construction, the flux of the $D=11$ four-form vanishes on $Y_9$, so these solutions describe the near horizon limit of $N$ M2-branes wrapping $M_2$.

The resulting geometries are again the class of GK geometries introduced in \cite{Kim:2006qu,Gauntlett:2007ts} and studied in \cite{Couzens:2018wnk, Boido:2022mbe}. More specifically, the existence of a Killing spinor guarantees the existence of an R-symmetry vector $\xi$, and it is possible to introduce a functional $\mathscr{S}$ of the supergravity fields on $Y_9$ with the following properties. Upon imposing the supersymmetry conditions, which do \textit{not} imply all the equations of motions for the fields, it is a function of $\xi$, and further extremizing the functional over the choice of $\xi$ is implied by imposing the remaining equations of motion, and $\mathscr{S}$ agrees with the area of $M_2$, and thus is related to the Bekenstein--Hawking entropy of the black hole with near horizon geometry $AdS_2\times M_2$.

It is also possible to introduce a further change of variables, in order to match 
other expressions in the literature.
We can introduce $\phi^I$ defined by
\begin{equation}
	\newc^I_\pm = \phi^I \pm b_0\, \Pnew^I \, ,
\end{equation}
which by construction are constrained by
\begin{equation}
	\sum_{I=0}^3 \phi^I = 2 - b_0 \frac{n_- - \sigma n_+}{n_- n_+} \, .
\end{equation}
Upon substitution of these variables in \eqref{eq:4d_STU_SBHOffShell}, the resulting expression matches (5.41) in \cite{Boido:2022mbe}, and thus the conjectures in \cite{Hosseini:2021fge, Faedo:2021nub}.

As discussed in section \ref{sec:UpliftAdS3} for uplifts of five-dimensional solutions, there is \textit{a priori} no reason for the correspondence of the gauged supergravity functional $S_{\rm BH}$ and the internal space one $\mathscr{S}$. Nonetheless, the geometric discussion surrounding \eqref{eq:EquivariantChernClasses} works in the same way, \textit{mutatis mutandis}, so again we can identify $\newc^I_p$, the values of the zero-degree components of the equivariant first Chern classes at a fixed point, with the weights of the circle action. Since in eleven dimensions the circle action on $M_2$ uplifts to the R-symmetry action on $Y_9$, we can identify $\newc^I_p$ with the R-symmetry charges.

\section{\texorpdfstring{$D=4$}{D=4} STU model with a hypermultiplet}\label{sec:sugramodel}

The model we will consider was given in \cite{Suh:2022pkg}. It is a subtruncation of the $U(1)^2 \subset SU(3)\subset SO(6)\subset SO(8)$ invariant consistent truncation of $D=4$ maximal $SO(8)$ gauged supergravity of \cite{Bobev:2018uxk}. The latter is an $\mathcal{N}=2$ gauged supergravity theory with three vector multiplets and one hypermultiplet and we consistently truncate one of the two complex scalar fields in the hypermultiplet to zero, and further
only considering solutions with $F^{{I}}\wedge F^{{J}}=0$ we can take the three complex scalars in the vector multiplets to be real \cite{Suh:2022pkg}. The model extends the $U(1)^4$ STU model of section~\ref{sec:4dSTU} with the addition of an extra complex scalar field $\rho\, 
\ex^{\ii\theta}$.

We use the results of \cite{Suh:2022pkg}, correcting the typos noticed in \cite{toappear}.
The bosonic part of the Lagrangian, in a {mostly plus} signature, is given by\footnote{Starting from \cite{toappear}, we have taken $\lambda_i\to \frac{1}{2}\varphi^i$,
$\varphi\to \frac{1}{2}\rho$, $A^I\to-\frac{1}{\sqrt 2} A^I$, and $g\to \frac{1}{\sqrt 2}$.}
\begin{align} \label{mlag}
\mathcal{L}=\frac{1}{16\pi G_{4}}\sqrt{-g}\Big[R&-\frac{1}{2}\sum_{i=1}^3(\partial\varphi^i)^2-\frac{1}{4}\sum_{I=0}^3(X^I)^{-2}F_{\mu\nu}^IF^{I\mu\nu}
-\mathcal{V} \notag \\
&-\frac{1}{2}(\partial\rho)^2-\frac{1}{2}\sinh^2\rho (D\theta)^2\Big]\,,
\end{align}
where $X^I(\varphi^i)$, $I=0,1,2,3$ are given in \eqref{eq:4d_GaugeFixingSymplecticCondition}.
In addition, we have 
\begin{align}\label{dthetastext}
 D\theta\equiv \dd\theta+ \sum_{I=0}^3 \zeta_I A^I\,,
\end{align}
where $\zeta_I=\frac{1}{2}(1,-1,-1,-1)$. The scalar potential $\mathcal{V}$ is given by
\begin{equation}
\label{eq:4d_HyperSTU_Potential}
\mathcal{V}={2}\left(\frac{\partial \mc{W} }{\partial\rho}\right)^2+2\sum_{i=1}^3\left(\frac{\partial {\mc{W}} }{\partial\varphi^i}\right)^2-\frac{3}{2} {\mc{W}}^2\,,
\end{equation}
where ${\mc{W}}$ is {the real superpotential} defined by
\begin{align}
\label{superpottext}
{\mc{W}} &=-\frac{1}{2}\sum_{I=0}^3 X^I+\zeta_IX^I\sinh^2\frac{\rho}{2}\,.
\end{align}

A solution preserves some of the supersymmetry of the maximal gauged supergravity theory if we can solve
\begin{align}\label{4d_susy1}
\Big[\nabla_\mu-\frac{\ii}{2}Q_\mu-\frac{1}{4} \mc{W} \Gamma_\mu+\frac{\ii}{16}\sum_{I=0}^3(X^I)^{-1}F_{\nu\rho}^{I}\Gamma^{\nu\rho}\Gamma_\mu\Big]\epsilon&=0\,,\nn\\
\Big[\Gamma^\mu\partial_\mu\varphi^i+2\partial_{\varphi^i}\mc{W}+\frac{\ii}{2}\sum_{I=0}^3\partial_{\varphi^i}(X^I)^{-1}{F}_{\mu\nu}^{I}\Gamma^{\mu\nu}\Big]\epsilon&=0\,, \notag \\
\Big[\Gamma^\mu\partial_\mu\rho+2\partial_\rho\mc{W}+{2\ii}\partial_\rho{Q}_\mu\Gamma^\mu\Big]\epsilon&=0\,,
\end{align}
where $\epsilon$ is a complex $D=4$ Dirac spinor and 
\begin{align} \label{hhbbdef}
Q_\mu&\equiv\,\frac{1}{2}\sum_{I=0}^3 A^I_\mu-\frac{1}{2}\left(\cosh\rho-1\right)D_\mu\theta\,.
\end{align}

Since this model extends the STU model, it admits the maximally supersymmetric $AdS_4$ vacuum solution discussed around \eqref{eq:4d_G4_Vacuum}, with vanishing matter fields and the $AdS_4$
having radius squared equal to $R^2_{\text{ABJM}}=1$. This solution uplifts to the $AdS_4\times S^7$ solution dual to ABJM theory. The model also admits a supersymmetric $AdS_4$ solution \cite{Warner:1983vz} with 
$R^2_{\text{mABJM}}\equiv \frac{4}{3\sqrt{3}}$, $\ex^{\frac{1}{2}\varphi^i}=3^{1/4}$ and $\tanh\frac{\rho}{2}=\frac{1}{\sqrt{3}}$
and vanishing gauge fields.
This solution preserves $SU(3)\times U(1)_R$ global symmetry and after uplifting to $D=11$ on $S^7$ is dual
to the $d=3$, $\mathcal{N}=2$ mABJM SCFT that arises as the IR fixed point in the RG flow of mass-deformed ABJM theory \cite{Ahn:2000aq,Corrado:2001nv,Benna:2008zy,Klebanov:2008vq}. 

The action is normalized with the $D=4$ Newton constant as given in \eqref{eq:4d_G4_Vacuum}. 
The free energy of the ABJM 
theory on $S^3$ is given by $F_{S^3}=\frac{\pi R^2_{\text{ABJM}}}{2G_{4}}=\frac{\pi \sqrt{2}}{3}N^{3/2}$.
Similarly, the 
free energy of the mABJM theory on $S^3$ is given by $F_{S^3}=\frac{\pi R^2_{\text{mABJM}}}{2G_{4}}=\frac{4\sqrt{2}\pi}{9\sqrt{3}}N^{3/2}$.

\subsection{The \texorpdfstring{$AdS_2$}{AdS2} ansatz}
We look for $AdS_2\times M_2$ solutions with metric again of the form \eqref{eq:4d_Ansatz}, 
with the gauge fields $A^I$ and the scalars $\lambda$, $\varphi^i$, $\rho$, $\theta$ all defined just on $M_2$. 
We can obtain the equations of motion for all of the fields by extremizing a $D=2$ action obtained by
reducing the $D=4$ action, analogous to the discussion in section \ref{subsec:AdS2Ansatz}. 
If we further impose the trace of the $D=4$ Einstein equation or, equivalently, 
the $D=2$ equation of motion for $\lambda$, we obtain the partially off-shell action
\begin{equation}
\label{eq:4d_HyperSTU_SoffShell}
S_2 |_{\text{POS}}\,  = \int_{M_2} \bigg[ \ex^{4\lambda} \mathcal{V} \, \vol - \frac{1}{2}\sum_{I=0}^3 (X^{I})^{-2} F^I_{12} \, F^I \bigg] \, .
\end{equation}
As in section \ref{subsec:AdS2Ansatz}, we further notice that on-shell we have
\begin{equation}
\label{eq:4d_STUHyper_S2OS}
S_2|_{\text{OS}} \, = - 2\int_{M_2} \ex^{2\lambda} \, \vol \, ,
\end{equation}
which is proportional to the Bekenstein--Hawking entropy, which again takes the form \eqref{eq:4d_SBH}.

The $D=4$ Dirac spinor can be written as \eqref{eq:4d_SpinorAnsatz} exactly as in the STU model. The Killing spinors therefore take 
the same form and satisfy the same algebraic equations. Some analysis of the differential conditions is given in appendix \ref{subapp:4d_BilinearsHyper}.
In particular, as in the STU model we have
\begin{equation}\label{kvgenexpmabjm}
	\dd \xi^\flat = 2 \left( 1 + \ex^\lambda \mc{W} {P}{S^{-1}} \right) P \, \vol  \, .
\end{equation}
We also need the relation
\begin{align}
\frac{1}{2}(X^I)^{-1}F_{12}^I=-\ex^{2\lambda}\mc{W}P S^{-1}- 2 \ex^\lambda\,.
\end{align}

\subsection{Equivariantly closed forms and localization}
There are several equivariant polyforms, $\Phi$, that can be constructed on $M_2$, satisfying $\dd_\xi\Phi\equiv (\dd-\xi\hook) \Phi=0$.  
Associated with the four gauge fields, as in the STU model we have
\begin{align}
\label{eq:4d_STUHyper_PhiFI}
\Phi^{F^I}=F^I-X^I \ex^\lambda P\,.
\end{align}
A new feature compared with the STU case is that a specific linear combination of
the $\Phi^{F^I}$, corresponding to the specific $U(1)$ that the complex scalar transforms under,
is equivariantly exact:
\begin{align}\label{eqexactmabjm}
\dd_\xi D\theta=\zeta_I \Phi^{F^I}\,.
\end{align}

Associated with the action we have that
\begin{align}
\label{eq:4d_STUHyper_Phi}
	\Phi^S = \ex^{4\lambda} \mathcal{V} \, \vol- \frac{1}{2}\sum_{I=0}^3 (X^{(I)})^{-2} F_{12}^I \, F^I +  \ex^{3\lambda} \mc{W} S +
	 \frac{1}{2}\ex^\lambda P \sum_{I=0}^3 (X^{(I)})^{-1}F_{12}^I  \, ,
	\end{align}
is equivariantly closed. Unlike the STU case, with $\rho\ne 0$ this is no longer the sum of two equivariantly closed forms, due to the presence of charged matter in the Maxwell equation.
From \eqref{eq:4d_STUHyper_S2OS} and \eqref{eq:4d_G4_Vacuum}, we can define the off-shell entropy function analogously to \eqref{eq:4d_SBH_OffShell} as
\begin{align}
S_{\rm BH} = -\frac{N^{3/2}}{6\sqrt{2}}\int_{M_2} \Phi^S \,.
\end{align}

We now assume that the $M_2$ has the topology of a spindle, with Killing vector $\xi$. The localization procedure proceeds
in a similar way to the $D=5$ case with hypermultiplets discussed in section \ref{sec:d5sugramodelhypers}. 
The orbifold singularities of $M_2$ are given by two integers 
$n_\pm\ge 1$ and we write the Killing vector as $\xi=b_0\partial_\varphi$, with $\Delta\varphi=2\pi$. We also write
$\epsilon_\pm = \mp b_0/n_\pm$ and 
{leave $P_+$ and $P_+/P_- \equiv  \sigma$ arbitrary for now.}

We set $S=1$ and define
\begin{align}\label{xprelhyperfour}
\newc^I\equiv  X^I \ex^\lambda P\,,
\end{align}
Using localization to obtain the fluxes, we get
\begin{align}\label{pifluxcpm}
\Pnew^I \equiv \frac{p^I}{n_+ n_-}\equiv	\frac{1}{4\pi}\int_{M_2} F^I= \frac{1}{2b_0}(\newc^I_+-\newc^I_-)  \, , \qquad p^I\in \mathbb{Z} \, .
\end{align}
Here we demand that the magnetic charges are suitably quantized, so that when the solutions are uplifted on $S^7$
we get a well-defined $AdS_2\times M_9$ solution of $D=11$ supergravity with $M_9$ an $S^7$ fibred over the spindle $M_2$. 
For the STU model, the factor of $1/(4\pi)$ in the quantization condition can be obtained\footnote{We also observe
that $\epsilon$ is a section of a Spin$^c$ bundle, as can be gleaned from its charge under the $A^I$ in \eqref{4d_susy1}, and our quantization condition is in agreement with general analysis for supersymmetric spindles in \cite{Ferrero:2021etw}.}
using the uplifting formulae in \cite{Cvetic:1999xp}. It is also important to emphasize that for solutions in which the charged scalar in the 
hypermultiplet is non-zero, we break a $U(1)\subset U(1)^4$ symmetry and correspondingly, we will lose a Killing vector on
the $S^7$ in the uplifted solution, but this does not alter the fact that we still need to impose a quantization condition on the fluxes to ensure the we have a good $S^7$ fibration over $M_2$, which is a topological constraint. 

Next, if we integrate \eqref{eqexactmabjm} over $M_2$ we deduce that the charges satisfy the constraint
\begin{align}\label{brokenu1mabjm}
\sum_{I=0}^3\zeta_I p^I=0\,.
\end{align}
From \eqref{dthetastext} we see that the magnetic flux associated with the $U(1)\subset U(1)^4$ for which the complex scalar is charged, must vanish.
Now if the complex scalar is non-vanishing at the poles, $\rho_\pm\ne0$, regularity implies that $D\theta|_\pm=0$
and then, considering the zero-form part of \eqref{eqexactmabjm}, we deduce the following
constraints on $\newc^I_\pm$:
\begin{align}\label{brokenabjmcconds}
\sum_{I=0}^3\zeta_I \newc^I_+=\sum_{I=0}^3\zeta_I \newc^I_-=0\,.
\end{align}
At the poles we therefore have $\zeta_I X^I\rvert_\pm=0$ and hence, from \eqref{superpottext}, we have
\begin{align}
\mc{W}|_\pm=-\frac{1}{2}\sum_{I=0}^3 X^I|_\pm\,.
\end{align}

Next, recalling that the weights $\epsilon_\pm$ of the action of the Killing vector at the poles are given by
$\dd\xi^\flat|_\pm=2\epsilon_\pm \vol_2$, using \eqref{kvgenexpmabjm} we deduce
\begin{align}\label{mabjmcpmsums}
\sum_{I=0}^3\newc^I_+= 2 + P_+ \frac{2b_0}{n_+} \,, \qquad
\sum_{I=0}^3\newc^I_-= 2 - \sigma P_+ \frac{2b_0}{n_-} \,.
\end{align}
Using the definition of the fluxes in \eqref{pifluxcpm}, we then deduce the following constraint on the R-symmetry flux
\begin{align}\label{rsymmabjm}
\sum_{I=0}^3 \Pnew^I= P_+ \frac{n_-+\sigma n_+}{n_-n_+}\,,\qquad
\Leftrightarrow  \qquad \sum_{I=0}^3 p^I = P_+ \left( n_- + \sigma n_+ \right) \, .
\end{align}

The localization of the action proceeds as in the $D=4$ STU model of section \ref{sec:4dSTU}
and we find
\begin{align}\label{smabjm}
\int_{M_2} \Phi^S = \frac{4\pi}{b_0} P_+ \left( \sqrt{\newc^0_+ \newc^1_+ \newc^2_+ \newc^3_+} -\sigma \sqrt{\newc^0_- \newc^1_- \newc^2_- \newc^3_- } \right) \, ,
\end{align}
and hence
\begin{align}
\label{eq:4d_STUHyper_SBHOffShell}
S_{\rm BH}= - P_+ \frac{\pi \sqrt{2} }{3}N^{3/2}\frac{1}{b_0} \left( \sqrt{\newc^0_+ \newc^1_+ \newc^2_+ \newc^3_+} -\sigma \sqrt{\newc^0_- \newc^1_- \newc^2_- \newc^3_- } \right) \,,
\end{align}
exactly the same as \eqref{eq:4d_STU_SBHOffShell}. Thus, the expression of the off-shell entropy function does not change upon including a hypermultiplet. Nonetheless, the extremization leads to a different result, since the constraints are different. 
The $\newc_\pm^I$ satisfy \eqref{pifluxcpm} and \eqref{mabjmcpmsums}, which implies the constraint on the R-symmetry flux
\eqref{rsymmabjm}, all as in the STU model. In addition we now need to impose \eqref{brokenabjmcconds} which imply 
the vanishing of the broken $U(1)$ flux in \eqref{brokenu1mabjm}. The
action depends on 9 variables $b_0, \newc^I_\pm$, for given $n_\pm,\sigma$. We can eliminate $\newc^I_-$ using \eqref{pifluxcpm}.
With the constraints on the R-symmetry flux in \eqref{rsymmabjm} and the vanishing of the flux associated
with the broken $U(1)$ in \eqref{brokenu1mabjm}, we are left with two independent fluxes
which we can take to be, for example, $p_1,p_2$.

One can now carry out an explicit extremization looking for solutions satisfying all of the relevant positivity conditions.
An example was presented in \cite{Suh:2022pkg} by studying the BPS equations directly, but it does not satisfy the
correct quantization conditions on the fluxes\footnote{In our language, if we take 
$(n_-,n_+)=(1,8)$ and fluxes $p^I=(-\frac{7}{2},-\frac{1}{6}-\frac{1}{\sqrt{3}},
-\frac{13}{6}-\frac{1}{\sqrt{3}},-\frac{7}{6}+\frac{2}{\sqrt{3}})$, we precisely recover the numerical values for the entropy 
just below (3.51) in \cite{Suh:2022pkg}, as well as the values for the scalar fields at the poles and $k$ given in (3.47) of \cite{Suh:2022pkg}, with $b_0=1/k$.}.

For the anti-twist class, we have found various explicit solutions that satisfy the appropriate positivity constraints, including $S_{BH}>0$.
We choose $P_+=-1$, as we did for the STU model, and summarize some examples in
table \ref{table1}, where we give the values of $ (n_+,n_-)$ and the magnetic fluxes $(p^1,p^2)$ with $(p^0,p^3)$ determined via
\begin{align}\label{p0p3hyper}
p^0=-\frac{1}{2}(n_--n_+)\,,\qquad p^3=p^0-p^1-p^2\,.
\end{align}
With $P_\pm=\mp 1$ from \eqref{xprelhyperfour} we have $x_+<0$ and $x_->0$. From
\eqref{mabjmcpmsums}, with $n_\pm\ge 1$ we deduce $n_->b_0>n_+$ and in turn, from
\eqref{pifluxcpm} that $p^I<0$.
We highlight that there is a symmetry that permutes $p_1,p_2,p_3$ and this can be utilized to obtain other cases.
For example, for $ (n_-,n_+)=(9,1)$ we can have 
$p^I=-(4,1,1,2)$ and $p^I=-(4,1,2,1)$ and $p^I=-(4,2,1,1)$.
We expect that for all these cases supergravity solutions exist: 
for the special cases with $p^1=p^2=-\frac{1}{6}(n_--n_+)$, we have explicit solutions of minimal $\mathcal{N}=2$ gauged supergravity and
for remaining cases one could check by numerically solving the BPS equations as in \cite{Suh:2022pkg}.  
\begin{table}[h!]
\footnotesize
\begin{center}
\begin{tabular}{|c|l|}
\hline
$(n_-,n_+)$ & $\qquad\qquad\qquad\quad$ Value of $-(p_1$, $p_2)$ \\\hline
(7,1)& (1,1)*\\\hline
(9,1)&(1,1)\\\hline
(11,1)&(1,1); \quad (2,2)\\\hline
(13,1)&(1,1); \quad (1,2);\quad (2,2)*\\\hline
(15,1)&(1,1); \quad (1,3); \quad (1,4);\quad (1;5); \quad (2,2)\\\hline
(8,2)&(1,1)*\\\hline
(10,2)&(1,1) \\\hline
(9,3)&(1,1)*\\\hline
\end{tabular}
\caption{Some values for spindle data in the anti-twist class with quantized magnetic fluxes $p^I$.
Cases with $p^1=p^2=- \frac{1}{6}(n_--n_+)$, marked with $*$, are associated with explicit solutions in minimal $\mathcal{N}=2$ gauged supergravity. The values of $p^0,p^3$ are determined from \eqref{p0p3hyper}, and there are additional examples that can be obtained by permuting $(p^1,p^2,p^3)$.\label{table1}}
\end{center}
\end{table}
As an illustration, we can be more explicit for a particular example by considering
$(n_-,n_+)=(15,3)$ and $p^I=-(6,3,2,1)$, finding 
\begin{align}
b_0&=4.23685\, ,\nn\\
\newc^0_-&=0.717543\,,\quad
\newc^1_-=0.331002\,,\quad
\newc^2_-=0.249196\,,\quad
\newc^3_-=0.137345\,,\,\nn\\
\newc^0_+&=-0.412284\,,\quad
\newc^1_+=-0.233912\,,\quad
\newc^2_+=-0.127413\,,\quad
\newc^3_+=-0.0509594\,,\nn\\
\ex^{2\lambda}|_-&={0.0901606}\,,\quad
\ex^{2\lambda}|_+={0.0250233}\,,\quad
S_{\rm BH}=0.0402617N^{3/2}\,.
\end{align}
Interestingly, the integers $(n_-,n_+)$ in this example are not co-prime and moreover, we find that there
are no solutions, with properly quantized fluxes, if we consider the $\mathbb{Z}_3$ quotient of the spindle and
take $(n_-,n_+)=(5,1)$. The cases with $(n_-,n_+)=(8,2), (10,2)$ and $(9,3)$ in the table have the same feature.
In previous works on spindles this possibility seems to have been overlooked. 

We have made a numerical search for any possible solutions in the twist class but have not found any.
It is therefore natural to conjecture that the twist class is obstructed.

We can also consider taking the $b_0\to 0$ limit, and set $n_\pm = 1 = \sigma$ to recover the $S^2$ case. This leads to the same entropy function 
\eqref{BZentropy} as in the STU case, but with the 
additional  hypermultiplet constraints 
$\zeta_I\newc^I=\zeta_I \Pnew^I=0$, and 
 precisely matches the result in \cite{Bobev:2018uxk}.

\section{\texorpdfstring{$D=4$, $\mathcal{N}=2$}{D=4 N=2} ungauged supergravity}
\label{sec:D4UngaugedSUGRA}

Similar to the discussion in section \ref{sec:d5ungauged} regarding $D=5$ supergravity,
it is straightforward to set the FI gaugings to zero and derive results for $D=4$ ungauged supergravity coupled to an arbitrary number of vector multiplets. 

The key difference with the discussion of the $D=4$ gauged supergravity theory
in section \ref{sec:D4SUGRA} is that the scalar potential $\mc{V}$ in \eqref{eq:4d_ScalarPotential} vanishes. However,
we can still study the $AdS_2 \times M_2$ spacetime \eqref{eq:4d_Ansatz} with $M_2 \cong \mathbb{WCP}_{[n_+, n_-]}$, and most of the derivations in section \ref{subsec:4d_Generic_Localization} go through, with the equivariantly closed form $\Phi^{\vol}$ 
\eqref{eq:4d_EquivariantPolyforms} now
being identically zero and correspondingly with a simplified form $\Phi^S$ in \eqref{phisdefn4}. 

We now find the R-symmetry flux \eqref{eq:4d_RSymmetryFlux} vanishes, thus imposing the equation
\begin{equation}
	n_- P_+ + n_+ P_- = 0 \, .
\end{equation}
With $n_\pm\ge 1$, we have $\sigma=P_+ P_-=-1$ and without loss of generality, this implies we can take $P_+ = -1$, $P_- = 1$ and $n_+ = n_- = n$. Thus, we deduce the new result
that ungauged supergravity does not admit any $AdS_2 \times M_2$ solutions with $M_2$ a spindle.
Furthermore, as in section \ref{sec:d5ungauged}, by going to the covering space we find that $M_2$ can always be taken to be $S^2$ with $n=1$. The constraints on $\newc^I$ \eqref{eq:4d_Constraint_BV} lead to $b_0=2$, so that $\xi$ rotates $S^2$ with weight 2. The off-shell entropy is
\begin{equation}
	S_{\rm BH} = \frac{\pi}{8G_4} \left[\ii \mc{F}( \newc^I_+) + \ii \mc{F}(\newc^I_-) \right] \, ,
\end{equation}
and we can introduce constants $\kappa^I$ via
\begin{equation}
	\newc^I_+ = 2 \left( \kappa^I + \Pnew^I \right) \, , \qquad \newc^I_- =  2 \left( \kappa^I - \Pnew^I \right) \, , 
\end{equation}
so that the extremization of $S_{\rm BH}$ has to be done over these $\kappa^I$, obtaining the equation
\begin{equation}
	\partial_J \mc{F}(\newc^I_+) =- \partial_J \mc{F}(\newc^I_-) \, .
\end{equation}
This equation may be solved by taking $\kappa^I=0$. 
Similarly to the discussion in section \ref{sec:d5ungauged}, $\kappa^I=0$ necessarily holds  for solutions which are invariant under the $SO(3)$ isometry 
group acting on $S^2$.\footnote{Also as in section \ref{sec:d5ungauged} we 
leave investigation of possible additional solutions for future work.}
With this solution, we then find
\begin{equation}\label{SBHungauged}
	S_{\rm BH} = \frac{\pi}{G_4} \ii \mc{F}(\Pnew^I) \, .
\end{equation}

At the poles, for these $AdS_2\times S^2$ solutions we find from the definition of $\newc^I$ that 
\begin{equation}
	 - \sqrt{2} \ex^{\cK/2}\ii X^I \rvert_\pm = \mp \left( \kappa^I \pm \Pnew^I \right) \ex^{-\lambda} \rvert_\pm\,,
\end{equation}
and for the case of $\kappa^I=0$, the constraint \eqref{eq:4d_eKFX_Constraint_2} gives
\begin{equation}\label{eKX}
	 \ex^{\cK/2}\ii X^I \rvert_\pm = - \frac{ \Pnew^I }{2 \sqrt{\ii \mc{F}( \pm \Pnew^I )}} \, .
\end{equation}
We immediately see one of the conclusions of the attractor mechanism (see {\it e.g.} \cite{Ferrara:1995ih, Behrndt:1996jn}): the value of the scalar fields at the horizon only depends on the charges, and not on the values at infinity. 
Our simple formula for the black hole entropy \eqref{SBHungauged} in the
purely magnetically charged case we have studied does not appear in \cite{Ferrara:1995ih, Behrndt:1996jn}, 
but it is straightforward to see that it agrees with those results. 
From equation (3.2) of \cite{Behrndt:1996jn} we have (restoring 
the Newton constant in their formula)
\begin{align}\label{Zhor}
S_{\rm BH} = \frac{\pi}{G_4} |Z_{\mathrm{hor}}|^2\, ,
\end{align}
where $Z$ is the central charge function, which  is evaluated 
at the horizon in \eqref{Zhor}. On the other hand 
equation (2.13) of  \cite{Behrndt:1996jn} gives the general formula
 (setting the electric charges to zero)
\begin{align}
|Z|^2 = \ex^{\mc{K}}|\Pnew ^I \mc{F}_I(X^J)|^2\, ,
\end{align}
where recall $\mc{F}_I \equiv \partial_{X^I}\mathcal{F}$.  
From \eqref{eKX} we may write $X^I = \gamma \Pnew^I$, 
where the proportionality constant $\gamma$ in 
fact drops out of both sides of \eqref{eKX}, 
using $\ex^{\mc{K}}\ii \mathcal{F}(X^I) = -\frac{1}{4}$ together 
with the fact that $\mc{F}$ is homogeneous degree 2.  
We then compute  $\gamma \Pnew^I \mc{F}_I (X^J) = 2\mc{F}(X^J)$ using 
homogeneity and Euler's formula, where we are evaluating at the extremal point 
$X^J = \gamma \Pnew^J$. 
Combing these formulae then gives
\begin{align}
S_{\rm BH} = \frac{\pi}{G_4} \ex^{\mc{K}} \cdot \left|\frac{2}{\gamma}\mathcal{F}(X^I)\right|^2 = 
\frac{\pi}{G_4} \frac{\ii\mc{F}(X^I)}{\gamma^2} = \frac{\pi}{G_4} \ii\mc{F}(\Pnew^I)\, ,
\end{align}
which agrees with  \eqref{SBHungauged}.

\section{Discussion}
\label{sec:Discussion}

Building on 
\cite{BenettiGenolini:2023kxp,BenettiGenolini:2023yfe,BenettiGenolini:2023ndb} we have shown
that equivariant localization provides a powerful tool to study black holes within gauged and ungauged supergravity.
Using the BVAB theorem we have constructed off-shell entropy functions for magnetically charged black holes of
$\mathcal{N}=2$, $D=4$ supergravity theories that have $AdS_2\times M_2$ horizons. We have also obtained
analogous results for magnetically charged black strings and rings (when they exist) of
$\mathcal{N}=2$, $D=5$ supergravity theories that have $AdS_3\times M_2$ horizons.
This provides a new perspective and also extends the attractor mechanism in both gauged and ungauged supergravity.

In order to use the localization technology we require $M_2$ to have a Killing vector and so we considered 
$M_2$ to be a spindle, $S^2$ or $T^2$. Naively, one might expect our approach to be useful 
only when the Killing vector on $M_2$ is an R-symmetry Killing vector constructed as a spinor bilinear. 
This is certainly the case when $M_2$ is a spindle and for this class we obtained many new results including 
showing that there are no solutions within ungauged supergravity. For the case of $T^2$ the two Killing vectors do not arise
as Killing spinor bilinears and our approach does not immediately lead to any new results. For
the case of $AdS_2\times S^2$ or $AdS_3\times S^2$ solutions of gauged supergravity, it is also the case that
the Killing vectors on $S^2$ are not arising from Killing spinor bilinears; nevertheless by considering
off-shell solutions where we consider this to be the case, we are able to construct off-shell entropy and central charges by 
taking a suitable limit. 

We considered supergravity models with arbitrary number of vector multiplets and illustrated the formalism with some specific examples. We studied the STU model for both $D=4$ and $D=5$ 
gauged supergravity and recovered results consistent with known explicit supergravity solutions. The solutions of the STU models are of particular interest as they can be uplifted on $S^7$ and $S^5$ to obtain exact solutions of $D=11$ and type IIB, respectively. The uplifted
solutions are of the form $AdS_2\times M_9$ and $AdS_3\times M_7$, respectively, with $M_7,M_9$ admitting a GK geometry.
We showed that our off-shell entropy functions and off-shell central charge are precisely equivalent to the gravitational block formulae
derived in
\cite{Boido:2022iye,Boido:2022mbe} using a very different approach in GK geometry. 

We also considered the STU model extended with a complex scalar arising from a hypermultiplet in 
both $D=4$ and $D=5$. We showed that the off-shell entropy and central charge functions that
one needs to extremize are the same as in the STU model, but the presence of the hypermultiplets imposes additional
constraints on the variables, as well as restricting the allowed magnetic fluxes. For the case when $M_2$ is a spindle
we illuminated and clarified the results of \cite{Arav:2022lzo,Suh:2022pkg} which have been obtained 
by studying the BPS equations directly. 
It would be interesting to study other models that can also be uplifted to string/M-theory. For example, one should be able to make contact with the solutions discussed in \cite{Halmagyi:2013sla} and \cite{Hristov:2023rel} with $S^2$ and spindle horizons, respectively.
More generally, it would certainly be of interest to extend our analysis to supergravity theories with arbitrary hypermultiplets.

In this paper we have discussed black objects that are magnetically charged, but we fully expect that the technology will generalize to include electric charge and non-trivial rotation. We have also just discussed 
two-derivative supergravity, and it would be interesting to extend our analysis to include higher derivative corrections and further investigate the ideas presented in \cite{Hristov:2021qsw}.

\section*{Acknowledgements}

\noindent 
We thank Seyed Morteza Hosseini, David Katona, Gabriel Lopes Cardoso, James Lucietti and Minwoo Suh for helpful discussions.
This work was supported in part by STFC grants  ST/T000791/1 and 
ST/T000864/1, and EPSRC grant EP/R014604/1. 
JPG is supported as a Visiting Fellow at the Perimeter Institute. 
PBG is supported by the SNSF Ambizione grant PZ00P2\_208666.
AL is supported by the Palmer scholarship in Mathematical Physics of Merton college.
JPG and JFS would like to thank the Isaac Newton Institute for Mathematical Sciences, Cambridge, for support and hospitality during the programme
``Black holes: bridges between number theory and holographic quantum information"
where work on this paper was undertaken. 

\appendix

\section{Reduction of \texorpdfstring{$D=5$}{D=5} Killing spinors}\label{app:5d}

In this appendix we reduce the $D=5$ Killing spinor equations 
\eqref{5dKSE} and \eqref{5d_susy1} via the $AdS_3\times M_2$
ansatz \eqref{ansatz5d}, 
and use these to deduce various algebraic and differential equations for bilinears in the Killing spinor $\zeta$ on $M_2$. 

\subsection{Killing spinor equations}\label{app:5dKSE}

The decomposition of the spinor $\epsilon$ and Cliff$(1,4)$ matrices 
is 
\begin{equation}\label{gamma5d}
	\epsilon = \vartheta\otimes\ex^{\lambda/2}\zeta\,,\quad
	\Gamma_\mathtt{i} = \beta_\mathtt{i}\otimes\gamma_3\,, \quad \Gamma_{a+2}=\mathbbm{1}\otimes\gamma_a\,, 
\end{equation}
where we denote frame indices on $AdS_3$ by $\mathtt{i}=0,1,2$ and those on $M_2$ by $a=1,2$. Here $\beta_{\mathtt{i}}$ generate Cliff$(1,2)$ and 
$\gamma_a$ generate Cliff$(2)$, with $\gamma_3=-\ii \gamma_{1}\gamma_2$.
The Killing spinor $\vartheta$ on $AdS_3$ satisfies
\begin{align}
	\nabla_\mathtt{i}\vartheta = \frac{1}{2}\beta_{\mathtt{i}}\vartheta\, .
\end{align}
Inserting this ansatz into \eqref{5dKSE} leads to the reduced 
spinor equations
\begin{align}\label{5d2dKSE}
	\nabla_a\zeta & = \left[\tfrac{1}{2}\left(1-\ex^{-\lambda}G_{IJ}X^I F_{12}^J\right)\gamma_a\gamma_3+\tfrac{\ii}{2}Q_a\right]\zeta\,,\nonumber\\
	0 & =\left[\slashed\del\lambda+\tfrac{1}{3}\ex^\lambda W+\left(1-\tfrac{1}{3}\ex^{-\lambda}G_{IJ}X^I F_{12}^J\right)\gamma_3\right]\zeta\,, \nonumber\\
	0 & =\left[\mathcal{G}_{ij}\slashed\del\varphi^j -\ex^\lambda\del_i W +\ex^{-\lambda}G_{IJ}\del_i X^I F_{12}^J\gamma_3\right]\zeta\, .
\end{align}
Using these one can derive the following useful relations, 
where $\mathbb{A}$ is any element of Cliff$(2)$:
\begin{align}\label{5dKSErel}
	\nabla_a(\zeta^\dagger\mathbb{A}\zeta)  & = \tfrac{1}{2}\left(1-\ex^{-\lambda}G_{IJ}X^I F_{12}^J\right)\zeta^\dagger[\mathbb{A},\gamma_a\gamma_3]_-\zeta\,,\nonumber\\
	0 & =(\del_a\lambda)\zeta^\dagger[\mathbb{A},\gamma^a]_\pm\zeta+\tfrac{1}{3}(1\pm 1)\ex^\lambda W \zeta^\dagger\mathbb{A}\zeta \nonumber\\
	& \quad+\left(1-\tfrac{1}{3}\ex^{-\lambda}G_{IJ}X^I F_{12}^J\right)\zeta^\dagger[\mathbb{A},\gamma_3]_\pm\zeta\,,\nonumber \\
	0 & =\mathcal{G}_{ij}(\del_a\varphi^j)\zeta^\dagger[\mathbb{A},\gamma^a]_\pm\zeta-(1\pm 1)\ex^\lambda \del_i W\zeta^\dagger\mathbb{A}\zeta \nonumber\\
	& \quad +\ex^{-\lambda}G_{IJ}\del_i X^I F_{12}^J\, \zeta^\dagger[\mathbb{A},\gamma_3]_\pm\zeta\, ,
\end{align}
where $[\cdot,\cdot]_\pm$ denote anti-commutator and commutator, respectively. 

\subsection{Bilinear equations}\label{app:5dbilinears}

Given the bilinear definitions \eqref{5dbilinears} we begin by 
noting the following identities:
\begin{align}
	\xi\hook\vol = -K\,, \quad F_{12}^I K=-\xi\hook F^I\,.
\end{align}
The relations in \eqref{5dKSErel} then lead to the following 
differential equations
\begin{align}\label{5dbilineareqns}
	&\dd S  =0 \,,\quad \diff K = 0 \, ,  \quad \dd \xi^\flat  = -2\left(2+\ex^{\lambda}WPS^{-1}\right)P\, \vol\,, \nonumber\\
	&\dd (\ex^\lambda P)  = - \frac{2}{3}G_{JK}X^J (\xi\hook F^K)\,,
	\quad 
	\dd (\ex^{4\lambda}S) = \frac{4}{3}\ex^{5\lambda}W  (\xi \hook \vol) \, ,\nonumber\\
	&\dd X^I  = -\ex^{-\lambda}\mathcal{G}^{ij}G_{JK} \del_i X^I\del_j X^J (\xi\hook F^K)P^{-1} \nn \\ & \qquad= -\ex^\lambda \mathcal{G}^{ij} \del_i X^I\del_j W (\xi \hook \vol)S^{-1}\, ,       
\end{align}
and algebraic equation
\begin{align}\label{5dalg}
	\ex^{-\lambda}G_{IJ}X^I F_{12}^J=3+\ex^{\lambda}WPS^{-1}\, .
\end{align}
Using equation \eqref{5dbilineareqns} we compute
\begin{align}
	\dd(X^I \ex^\lambda P)&=(\dd X^I)\ex^\lambda P + X^I \dd(\ex^\lambda P)
	\nonumber \\
	&=-G_{JK}\Big(\mathcal{G}^{ij}\del_i X^I\del_j X^J+\frac{2}{3}X^I X^J\Big)(\xi\hook F^K)\nonumber\\
	& =-G_{JK}G^{IJ}(\xi\hook F^K)=-\xi\hook F^I\,,
\end{align}
where we have used 
\eqref{GIJ}. This proves that 
the multi-form $\Phi^{F^I}=F^I- X^I \ex^\lambda P$ is equivariantly closed. 

Next we similarly compute 
\begin{align}\label{closedphivol}
	\dd(\ex^{4\lambda}W S)&=\xi_I(\dd X^I)\ex^{4\lambda} S +W \dd(\ex^{4\lambda} S)\nonumber \\
	&=-\ex^{5\lambda}\Big(\mathcal{G}^{ij}\del_i W\del_jW -\frac{4}{3}W^2\Big)(\xi\hook\vol)\nonumber \\
	& =-\ex^{5\lambda}\mathcal{V}\, (\xi\hook\vol)\,,
\end{align}
where we have used \eqref{P5d}. This proves that 
the multi-form $\Phi^\vol=\ex^{5\lambda}\mathcal{V}\, \vol-\ex^{4\lambda}W S$ is equivariantly closed. 

\subsection{STU with hypermultiplet}

We now turn to the STU model with hypermultiplets considered in section \ref{sec:d5sugramodelhypers}.
Compared to the general vector model presented in the previous subsection, this amounts to setting  $\xi_I=(1,1,1)$ and considering an additional complex scalar field $\rho\, \ex^{\ii\theta}$. Every equation from the previous section follows through directly, except for the last relation \eqref{closedphivol}. There are also some additional constraints for $\rho$. 
Note that with the STU prepotential $\mathcal{F}=X^1 X^2 X^3$, the metrics defined in \eqref{metrics} are simply 
\begin{align}
	G_{IJ}=\frac{1}{2}\begin{pmatrix}(X^1)^{-2}& 0 & 0 \\ 0 &(X^2)^{-2}& 0 \\ 0 & 0 &(X^3)^{-2}\end{pmatrix}\,, \quad \mathcal{G}_{ij}=\frac{1}{2}\delta_{ij}\,.
\end{align}

Under the ans\"atze \eqref{ansatz5d} and \eqref{gamma5d} the Killing spinor equations \eqref{5d_susy1} reduce to 
\begin{align}
	\nabla_a\zeta & = \Big[\tfrac{1}{2}\Big(1-\tfrac{1}{2}\ex^{-\lambda}\sum_I(X^I)^{-1} F_{12}^I\Big)\gamma_a\gamma_3+\tfrac{\ii}{2}Q_a\Big]\zeta\,,\nonumber\\
	0 & =\Big[\slashed\del\lambda+\tfrac{1}{3}\ex^\lambda W+\Big(1-\tfrac{1}{6}\ex^{-\lambda}\sum_I(X^I)^{-1} F_{12}^I\Big)\gamma_3\Big]\zeta\,, \nonumber\\
	0 & =\Big[\slashed\del\varphi^i -2\ex^\lambda\del_i W -\ex^{-\lambda}\sum_I \partial_i(X^I)^{-1} F_{12}^I\gamma_3\Big]\zeta\, \nn \\
	0&=[\slashed{\del}\vpvar-2\ex^{\lambda}\del_\vpvar W+4\ii(\del_\rho Q_a)\gamma^a]\zeta\,,
\end{align}
which are the same equations as \eqref{5d2dKSE}, but where the last line is new. Again these
give the constraints \eqref{5dKSErel} plus the following relation for $\rho$ where $\mathbb{A}\in\mathrm{Cliff}(2)$
\begin{align}\label{5dKSErelhyper}
	0= \, &(\del_a\vpvar)\zeta^\dagger[\mathbb{A},\gamma^a]_\pm\zeta-2(1\pm1)\ex^{\lambda}\del_\vpvar W\zeta^\dagger\mathbb{A}\zeta+4\ii(\del_\vpvar Q_a)\zeta^\dagger[\mathbb{A},\gamma^a]_\mp\zeta\,.
\end{align}
From these we deduce the same differential constraints as before, plus a new one on the last line:
\begin{align}
	&\dd S  =0 \,,\quad \diff K = 0 \, ,  \quad \dd \xi^\flat  = -2\left(2+\ex^{\lambda}WPS^{-1}\right)P\, \vol\,, \nonumber\\
	&\dd (\ex^\lambda P)  = - \frac{1}{3}\sum_I(X^I)^{-1} (\xi\hook F^I)\,,
	\quad 
	\dd (\ex^{4\lambda}S) = \frac{4}{3}\ex^{5\lambda}W  (\xi \hook \vol) \, ,\nonumber\\
	&\dd X^I  = -\sum_{i,J}\ex^{-\lambda} \del_i X^I\del_i (X^J)^{-1} (\xi\hook F^J)P^{-1}\, , \nn \\
	&\dd \varphi^i=-2\ex^{\lambda}\del_i W (\xi\hook\vol)S^{-1}\, , \nn \\
	&\dd \vpvar=-2\ex^{\lambda}\del_\vpvar W (\xi\hook\vol)S^{-1}-4(*\del_\vpvar Q)PS^{-1}\,,
\end{align}
such that
\begin{align}\label{dW}
	\dd W = -2\ex^{\lambda}\Big[\sum_i(\del_i W)^2+(\del_\vpvar W)^2\Big](\xi\hook\vol)S^{-1}-4(\del_\vpvar W)(*\del_\vpvar Q)PS^{-1},
\end{align}
where
\begin{align}\label{dWdQ}
	\del_\vpvar W = \frac{1}{2}\zeta_I X^I\sinh\rho\,, \quad \del_\vpvar Q = - \frac{1}{4}D\theta\sinh\rho\,.
\end{align}
We also record the Maxwell equation
\begin{align}\label{MaxwellSTU}
	\dd\left[\ex^{\lambda}(X^I)^{-2}F_{12}^I\right]=\ex^{3\lambda}\zeta_I\sinh^2\rho\, (\star D\theta)\,.
\end{align}

As before $\Phi^{F^I}$ is equivariantly closed: 
\begin{align}
	\dd(X^I \ex^\lambda P)&=(\dd X^I)\ex^\lambda P + X^I \dd(\ex^\lambda P)
	\nonumber \\
	&=-\sum_{i,J}\Big[\del_i X^I\del_i (X^J)^{-1}+\frac{1}{3}X^I (X^J)^{-1}\Big](\xi\hook F^J)\nonumber\\
	& =-\sum_J\Big(\sum_i\varsigma_{Ii}\varsigma_{Ji}+\frac{1}{3}\Big)X^I(X^J)^{-1}(\xi\hook F^J)=-\xi\hook F^I\,,
\end{align}
where we have summarized \eqref{X5d} as
\begin{equation}
	X^I = \exp \Big( \sum_i  \varsigma_{Ii} \varphi^i \Big) \, ,\quad \Big(\sum_i\varsigma_{Ii}\varsigma_{Ji}+\frac{1}{3}\Big)=\delta_{IJ}\,.
\end{equation}

On the other hand using \eqref{dW} and the definition \eqref{P5dhpyer} of the scalar potential
\begin{align}
	\dd(\ex^{4\lambda}W S)
	=-\ex^{5\lambda}\mathcal{V}\, (\xi\hook\vol)-4\ex^{4\lambda}(\del_\vpvar W)(*\del_\vpvar Q)P\,,
\end{align}
we find that $\Phi^\vol=\ex^{5\lambda}\mathcal{V}\, \vol-\ex^{4\lambda}W S$ is not equivariantly closed. Instead using \eqref{dWdQ} and \eqref{MaxwellSTU}
we obtain that
\begin{align}
	\dd_\xi \Phi^\vol& = 4\ex^{4\lambda}(\del_\vpvar W)(*\del_\vpvar Q)P \nn \\& =-\frac{1}{2}\sum_I \dd\left[\ex^{\lambda}(X^I)^{-2}F_{12}^I\right] X^I\ex^{\lambda}P\,.
\end{align}
This proves that the multi-form having the action as the top form,
\begin{align}
	\Phi^S=\frac{2}{3}\Big[\Phi^\vol-\frac{1}{2}\sum_I\left(\ex^{\lambda}(X^I)^{-2}F_{12}^I\right)\Phi^{F^I}\Big]\,,
\end{align}
is equivariantly closed.

Finally we show that $\zeta_I\Phi^{F^I}$ is equivariantly exact. Indeed \eqref{5dKSErelhyper} gives   
\begin{align}
	\xi\hook(\del_\vpvar Q)=\tfrac{1}{2}\ex^{\lambda}(\del_\vpvar W)P \quad \implies \quad 
	\xi\hook D\theta=-\zeta_IX^I\ex^{\lambda}P\,,
\end{align}
where we used the explicit expressions \eqref{dWdQ}, so that 
\begin{align}
	\dd_\xi D\theta=-\zeta_I F^I +\zeta_IX^I\ex^{\lambda}P = -\zeta_I\Phi^{F^I}\,.
\end{align}

\section{Reduction of \texorpdfstring{$D=4$}{D=4} Killing spinors}
\label{app:4dBilinears}

In this appendix we reduce the $D=4$ Killing spinor equations \eqref{eq:4d_GravitinoVariation} and \eqref{eq:4d_GauginoVariation} via the ansatz \eqref{eq:4d_Ansatz}, \eqref{eq:4d_SpinorAnsatz}. This is analogous to the procedure described in appendix~\ref{app:5d}, and will lead to differential and algebraic equations for the bilinears in $\zeta$. Because of the similarities, we shall be brief.

One relevant difference with the five-dimensional case is the presence of the Abelian R-symmetry factor mentioned in footnote~\ref{footnote:Rsymmetry}. Concretely, this transformation acts on the fields by
\begin{equation}
\label{eq:app_U1RFields}
	\epsilon \to \ex^{ - \ii \frac{\alpha}{2} \Gamma_5} \epsilon \, , \qquad {\ex^{\mc{K}/2}} X^I \to \ex^{ - \ii \alpha } {\ex^{\mc{K}/2}} X^I \, , \qquad \mc{A}_\mu \to \mc{A} + \partial_\mu \alpha \, ,
\end{equation}
and under this transformation the equations \eqref{eq:4d_GravitinoVariation} and \eqref{eq:4d_GauginoVariation} transform covariantly, that is, schematically
\begin{equation}
\label{eq:app_U1RKSE}
	(\delta\psi_\mu, \delta \lambda) \left( \ex^{ - \ii \frac{\alpha}{2} \Gamma_5} \epsilon, \ex^{ - \ii \alpha } {\ex^{\mc{K}/2}}X^I, \mc{A} + \partial_\mu \alpha \right) = \ex^{ - \ii \frac{\alpha}{2} \Gamma_5} (\delta\psi_\mu, \delta \lambda) \left( \epsilon, {\ex^{\mc{K}/2}}X^I, \mc{A} \right) \, ,
\end{equation}
where $\psi_\mu$ and $\lambda$ are the gravitino and gaugino, respectively. We shall see in \ref{subapp:4d_BilinearsHyper} that this transformation allows us to change the reality of the sections.

\subsection{Killing spinor equations}

We decompose the spinor and Clifford algebra Cliff$(1,3)$ as
\begin{equation}
\label{eq:4d_SpinorAnsatz_app}
	\epsilon = \vartheta \otimes \ex^{\lambda/2} \zeta\, , \quad \Gamma_{\mathtt{i}} = \beta_{\mathtt{i}} \otimes \gamma_3 \, , \quad \Gamma_{a+1} = \mathbbm{1} \otimes \gamma_a \, ,
\end{equation}
where $\beta_{\mathtt{i}}$, with ${\mathtt{i}} = 0, 1$, generate Cliff$(1,1)$ with $\beta_3 \equiv - \beta_{01}$, and $\gamma_a$ generate Cliff$(2)$, with $\gamma_3 \equiv - \ii \gamma_{12}$. We also assume that 
\begin{equation}
\label{eq:AdS2_KS}
	\nabla_{\mathtt{i}} \vartheta = - \frac{1}{2} \beta_{\mathtt{i}} \vartheta \, .
\end{equation}
Inserting this ansatz in \eqref{eq:4d_GravitinoVariation} gives
\begin{align}
	0 &= \vartheta \otimes \left[ \nabla_a \zeta + \frac{1}{2}\gamma_a \slashed{\partial}\lambda \, \zeta - \frac{\ii}{4} \xi_I A^I_a \, \zeta + \frac{1}{2\sqrt{2}} \ex^\lambda \ex^{\cK/2} \Im W \, \gamma_a \zeta \right. \nn \\
	& \qquad \qquad  \left. + \frac{1}{2\sqrt{2}} \ex^{-\lambda} \Im \mc{N}_{IJ} F^J_{12} \ex^{\cK/2} \Im X^I \, \gamma_3 \gamma_a \zeta \right] \nn \\
 &  \ \ \ + \frac{\ii}{2} \beta_3 \vartheta \otimes \left( \mc{A}_a \gamma_3\zeta + \frac{1}{\sqrt{2}} \ex^\lambda \ex^{\cK/2} \Re W \, \gamma_a \gamma_3 \zeta \right. \nn \\
\label{eq:app_AnsatzGravitino_1}
 & \qquad \qquad \qquad \left. - \frac{1}{\sqrt{2}} \ex^{-\lambda} \Im \mc{N}_{IJ} F^J_{12} \ex^{\cK/2} \Re X^I \, \gamma_a \zeta \right) \, , \\[5pt]
	0 &= \frac{1}{2} \beta_\lambda \vartheta \otimes \left[ - \zeta + \gamma_3 \slashed{\partial}\zeta + \frac{1}{\sqrt{2}} \ex^\lambda \ex^{\cK/2} \Im W \, \gamma_3 \zeta + \frac{1}{\sqrt{2}} \ex^{-\lambda} \Im \mc{N}_{IJ} F^J_{12} \ex^{\cK/2} \Im X^I \, \zeta \right] \nn \\
\label{eq:app_AnsatzGravitino_2}
	& \ \ \ + \frac{\ii}{2} \ex^{\lambda} \beta_\lambda \beta_3 \vartheta \otimes \ex^{\cK/2} \left[ \frac{1}{\sqrt{2}} \Re W \, \zeta + \frac{1}{\sqrt{2}} \ex^{-2\lambda} \Im \mc{N}_{IJ} F^J_{12} \Re X^I \gamma_3 \zeta \right] \, .
\end{align}
The gaugino variation \eqref{eq:4d_GauginoVariation}, instead, gives
\begin{align}
	0 &= \vartheta \otimes \left[ - \frac{1}{\sqrt{2}} \ex^{-2\lambda} \Im \mc{N}_{IJ} F^J_{12} \ex^{\cK/2} \Im (\mc{G}^{\bar{i} j} \nabla_j X^I) \gamma_3 \zeta + \ex^{-\lambda} \slashed{\partial}\Re z^i \, \zeta - \frac{1}{\sqrt{2}} \Im( \mc{G}^{\bar{i}j} \nabla_j W) \, \zeta \right] \nn \\
	& \ \ \ + \ii \beta_3 \vartheta \otimes \left[ - \frac{1}{\sqrt{2}} \ex^{-2\lambda} \Im \mc{N}_{IJ} F^J_{12} \ex^{\cK/2} \Re (\mc{G}^{\bar{i}j}\nabla_j X^I) \zeta - \ex^{-\lambda} \slashed{\partial} \Im z^i \, \gamma_3 \zeta \right. \nn \\
\label{eq:app_AnsatzGaugino}
	& \left. - \frac{1}{\sqrt{2}} \Re ( \mc{G}^{\bar{i}j}\nabla_j W) \gamma_3 \zeta \right] \, .
\end{align}
Motivated by known examples ({\it e.g.} \cite{Hristov:2010ri}), in order to solve the equations above we impose the following conditions
\begin{equation}
\label{eq:4d_Constraint_App}
	z^i \in \R \, , \qquad X^I \in \ii \R \, , 
\end{equation}
 and so for instance we may write $\Im W = - \ii W$. These conditions imply that the equations simplify to
\begin{align}
\label{eq:app_SUSYEqns_4d}
	0 &= \nabla_a \zeta - \frac{\ii}{4} \xi_I A^I_a \, \zeta + \left( - \frac{1}{2} - \frac{1}{\sqrt{2}} \ii \ex^{-\lambda} \ex^{\cK/2} \Im \mc{N}_{IJ} F^J_{12} X^I \right) \gamma_3 \gamma_a \zeta \, , \nn \\
	0 &= \slashed{\partial}\zeta - \frac{\ii}{\sqrt{2}} \ex^\lambda \ex^{\cK/2} W \, \zeta + \left( - 1 - \frac{1}{\sqrt{2}} \ii \ex^{-\lambda} \Im \mc{N}_{IJ} F^J_{12} \ex^{\cK/2} X^I \right) \gamma_3 \zeta \, , \nn \\
	0 &= \frac{1}{\sqrt{2}} \ii\ex^{-2\lambda}\Im \mc{N}_{IJ} F^J_{12} \ex^{\cK/2} \mc{G}^{ij} \nabla_j X^I \, \gamma_3 \zeta + \ex^{-\lambda} \slashed{\partial} z^i \, \zeta + \frac{\ii}{\sqrt{2}} \mc{G}^{ij}  \nabla_j W \, \zeta \, .
\end{align}
If $\mathbb{A}$ is any element of Cliff$(2)$, then the following relations hold
\begin{align}
	\nabla_a ( \zeta^\dagger \mathbb{A} \zeta ) &= \left( - \frac{1}{2} - \frac{1}{\sqrt{2}} \ii \ex^{-\lambda} \ex^{\cK/2} \Im \mc{N}_{IJ} F^J_{12} X^I \right) \zeta^\dagger [\gamma_3\gamma_a, \mathbb{A} ]_- \zeta \, , \nn \\[5pt]
	0 &= (\partial_a\lambda) \zeta^\dagger [\mathbb{A}, \gamma^a ]_\pm \zeta - \frac{1\pm 1}{\sqrt{2}} \ii \ex^\lambda \ex^{\cK/2} W \, \zeta^\dagger \mathbb{A} \zeta \nn \\
	& \ \ \ + \left( -1 - \frac{1}{\sqrt{2}} \ii \ex^{-\lambda} \Im \mc{N}_{IJ} F^J_{12} \ex^{\cK/2} X^I \right) \zeta^\dagger [\mathbb{A}, \gamma_3 ]_\pm \zeta \, , \nn \\[5pt]
	0 &= \ex^{-\lambda} \partial_a z^i \, \zeta^\dagger [\mathbb{A}, \gamma^a ]_\pm \zeta + \frac{1\pm 1}{\sqrt{2}} \ii \ex^{\cK/2} \mc{G}^{ij} \nabla_j W \, \zeta^\dagger \mathbb{A} \zeta \nn \\
	& \ \ \ + \frac{1}{\sqrt{2}} \ii \ex^{-2\lambda} \Im \mc{N}_{IJ} F^J_{12} \ex^{\cK/2} \mc{G}^{ij} \nabla_j X^I \zeta^\dagger [\mathbb{A}, \gamma_3 ]_\pm \zeta \, .
\end{align}

\subsection{Bilinear equations}

Using the definitions \eqref{5dbilinears} for the bilinears constructed using $\zeta$, we find the following differential relations
\begin{align}
\label{eq:app_SUSYBilinears_4d_1}
	\dd S &= 0 \, , \quad \dd( \ex^\lambda P) = \frac{1}{\sqrt{2}} \Im \mc{N}_{IJ} \ex^{\cK/2} \Im X^I \, \xi \hook F^J \, , \quad \dd K = 0 \, , \\
\label{eq:4d_dxib}
	\dd \xi^\flat &= 2 P \left( 1 + \sqrt{2} \ii P S^{-1} \ex^\lambda \ex^{\cK/2} W \right) \vol \, , \\
	\dd ( \ex^\lambda S ) &= - \frac{\ii}{\sqrt{2}}  \ex^{2\lambda} \ex^{\cK/2} W \, \xi \hook \vol \, , \\
	\dd ( \ex^{\cK/2} X^I ) &= \ex^{\cK/2} \nabla_i X^I \, \dd z^i = - \frac{1}{\sqrt{2}} \ii P^{-1} \ex^{-\lambda} \Im \mc{N}_{JL} \ex^{\cK} \mc{G}^{ij} \nabla_i X^I \nabla_j X^L \, \xi \hook F^J \nn \\
\label{eq:app_SUSYBilinears_4d_2}
	&= \frac{1}{\sqrt{2}} S^{-1}  \ii \ex^\lambda \ex^{\cK} \mc{G}^{ij} \nabla_i X^I \nabla_j W \, \xi \hook \vol \, ,
\end{align}
and the algebraic condition
\begin{equation}
\label{eq:4d_AlgebraicF12}
	- \frac{1}{\sqrt{2}} \ex^{-\lambda} \Im \mc{N}_{IJ}F^I_{12} \ex^{\mc{K}/2} \ii X^J = 1 + \frac{\ii}{\sqrt{2}} PS^{-1} \ex^\lambda \ex^{\mc{K}/2} W \, .
\end{equation}
This allows us to prove the following
\begin{align}
	\dd ( \ex^{\cK/2}\ii X^I \ex^\lambda P) &= \frac{1}{\sqrt{2}} \Im \mc{N}_{JL} \ex^{\cK} \left( \mc{G}^{ij} \nabla_i X^I \nabla_j X^L + X^I X^L \right) \, \xi \hook F^J \nn \\
	&= \frac{1}{2\sqrt{2}} \xi \hook F^I \, ,
\end{align}
where we used the identity
\begin{equation}
	\ex^{\cK} \left( \mc{G}^{i\bar{j}} \nabla_i X^I \nabla_{\bar{j}} \overline X^J + \overline{X}^I X^J \right) = - \frac{1}{2} [(\Im \mc{N})^{-1}]^{IJ} \, .
\end{equation}
This relation implies that $\Phi^{F^I}$ in \eqref{eq:4d_EquivariantPolyforms} is equivariantly closed.

Similarly
\begin{align}
	\dd \left( \sqrt{2} \ex^{3\lambda} \ex^{\cK/2} \ii W S \right) &= - \ex^{4\lambda} \ex^{\cK} \left( \mc{G}^{ij} \nabla_i W \nabla_j W - 3 W^2 \right) \, \xi \hook \vol \nn \\
	&= \ex^{4\lambda} \mc{V} \xi \hook \vol \, ,
\end{align}
using the definition \eqref{eq:4d_ScalarPotential_v2}. This relation, together with Maxwell's equations, implies that $\Phi^{\rm vol}$ in \eqref{eq:4d_EquivariantPolyforms} is equivariantly closed.

\subsection{STU with hypermultiplet}
\label{subapp:4d_BilinearsHyper}

In this appendix we present the bilinear equations and show that the polyform $\Phi^S$ in \eqref{eq:4d_STUHyper_Phi} is equivariantly closed. This model differs from the previous discussion of the theory with an arbitrary number of vector multiplets because there is an additional supersymmetry equation coming from the variation of the hyperino, and because of a difference choice of conventions (as mentioned below \eqref{eq:4d_GaugeFixingSymplecticCondition}, there is a $U(1)$ phase rotation that is required to go between
the two), so we repeat some steps for the reader's convenience.

First, we clarify the latter point for the STU model (that is, in absence of the hypermultiplet). As remarked earlier, the phase of the sections $X^I$ is immaterial, since the $U(1)$ transformation \eqref{eq:app_U1RFields} can be used to trade it for a change of phase for the Killing spinor according to \eqref{eq:app_U1RKSE}. Concretely, if instead of choosing the ansatz \eqref{eq:4d_SpinorAnsatz_app}, we choose the spinor $\epsilon' = \ex^{ - \ii \frac{\pi}{4} \Gamma_5} \epsilon$, then we find the following equations instead of \eqref{eq:app_AnsatzGravitino_1}, \eqref{eq:app_AnsatzGravitino_2} and \eqref{eq:app_AnsatzGaugino}
\begin{align}
	0 &= \vartheta \otimes \left[ \nabla_a \zeta + \frac{1}{2}\gamma_a \slashed{\partial}\lambda \, \zeta - \frac{\ii}{4} \xi_I A^I_a \, \zeta + \frac{1}{2\sqrt{2}} \ex^\lambda \ex^{\cK/2} \Re W \, \gamma_a \zeta \right. \nn \\
	& \qquad \qquad  \left. + \frac{1}{2\sqrt{2}} \ex^{-\lambda} \Im \mc{N}_{IJ} F^J_{12} \ex^{\cK/2} \Re X^I \, \gamma_3 \gamma_a \zeta \right] \nn \\
 &  \ \ \ + \frac{\ii}{2} \beta_3 \vartheta \otimes \left( \mc{A}_a \gamma_3\zeta - \frac{1}{\sqrt{2}} \ex^\lambda \ex^{\cK/2} \Im W \, \gamma_a \gamma_3 \zeta \right. \nn \\
 & \qquad \qquad \qquad \left. + \frac{1}{\sqrt{2}} \ex^{-\lambda} \Im \mc{N}_{IJ} F^J_{12} \ex^{\cK/2} \Im X^I \, \gamma_a \zeta \right) \, , \\[5pt]
	0 &= \frac{1}{2} \beta_\lambda \vartheta \otimes \left[ - \zeta + \gamma_3 \slashed{\partial}\zeta + \frac{1}{\sqrt{2}} \ex^\lambda \ex^{\cK/2} \Re W \, \gamma_3 \zeta + \frac{1}{\sqrt{2}} \ex^{-\lambda} \Im \mc{N}_{IJ} F^J_{12} \ex^{\cK/2} \Re X^I \, \zeta \right] \nn \\
	& \ \ \ + \frac{\ii}{2} \ex^{\lambda} \beta_\lambda \beta_3 \vartheta \otimes \ex^{\cK/2} \left[ - \frac{1}{\sqrt{2}} \Im W \, \zeta - \frac{1}{\sqrt{2}} \ex^{-2\lambda} \Im \mc{N}_{IJ} F^J_{12} \Im X^I \gamma_3 \zeta \right] \, , \\[5pt]
	0 &= \vartheta \otimes \left[ - \frac{1}{\sqrt{2}} \ex^{-2\lambda} \Im \mc{N}_{IJ} F^J_{12} \ex^{\cK/2} \Re (\mc{G}^{\bar{i} j} \nabla_j X^I) \gamma_3 \zeta + \ex^{-\lambda} \slashed{\partial}\Re z^i \, \zeta - \frac{1}{\sqrt{2}} \Re( \mc{G}^{\bar{i}j} \nabla_j W) \, \zeta \right] \nn \\
	& \ \ \ + \ii \beta_3 \vartheta \otimes \left[ \frac{1}{\sqrt{2}} \ex^{-2\lambda} \Im \mc{N}_{IJ} F^J_{12} \ex^{\cK/2} \Im (\mc{G}^{\bar{i}j}\nabla_j X^I) \zeta - \ex^{-\lambda} \slashed{\partial} \Im z^i \, \gamma_3 \zeta \right. \nn \\
	& \left. + \frac{1}{\sqrt{2}} \Im ( \mc{G}^{\bar{i}j}\nabla_j W) \gamma_3 \zeta \right] \, .
\end{align}
These are simplified by the choice $z^i \in \R$ and $X^I\in \R$ (compare with \eqref{eq:4d_Constraint_App}), assuming which we find
\begin{align}
	0 &= \nabla_a \zeta - \frac{\ii}{4} \xi_I A^I_a \, \zeta + \left( - \frac{1}{2} + \frac{1}{\sqrt{2}} \ex^{-\lambda} \ex^{\cK/2} \Im \mc{N}_{IJ} F^J_{12} X^I \right) \gamma_3 \gamma_a \zeta \, , \nn \\
	0 &= \slashed{\partial}\zeta +  \frac{1}{\sqrt{2}} \ex^\lambda \ex^{\cK/2} W \, \zeta - \left( 1 - \frac{1}{\sqrt{2}} \ex^{-\lambda} \Im \mc{N}_{IJ} F^J_{12} \ex^{\cK/2} X^I \right) \gamma_3 \zeta \, , \nn \\
\label{eq:app_SimplifiedEqs_2}
	0 &= - \frac{1}{\sqrt{2}} \ex^{-2\lambda}\Im \mc{N}_{IJ} F^J_{12} \ex^{\cK/2} \mc{G}^{\bar{i}j}\nabla_j X^I  \gamma_3 \zeta + \ex^{-\lambda} \slashed{\partial}\Re z^i \, \zeta - \frac{1}{\sqrt{2}} \mc{G}^{\bar{i}j} \nabla_j W \, \zeta \, .
\end{align}

\medskip

Moving now to the concrete case of the STU model with a hypermultiplet, the supersymmetry equations are \eqref{4d_susy1} and inserting the ans\"{a}tze \eqref{eq:4d_SpinorAnsatz_app}, \eqref{eq:AdS2_KS} leads to the equations
\begin{align}
	0 &= \nabla_a {\zeta} - \frac{\ii}{2} Q_a {\zeta} + \frac{1}{2} \left( -1 - \frac{1}{2} \ex^{-\lambda} \sum_I (X^I)^{-1} F^I_{12} \right) \gamma_3 \gamma_\alpha {\zeta} \, , \nn \\	
	0 &= \slashed{\partial} \lambda \, {\zeta} - \frac{1}{2} \mc{W} \ex^{\lambda} \, {\zeta} + \left( -1 - \frac{1}{4} \ex^{-\lambda} \sum_I (X^I)^{-1} F^I_{12}  \right) \gamma_3 {\zeta} \, , \nn \\
	0 &= - \ex^{-2\lambda} \sum_I \partial_{\varphi^i} (X^I)^{-1} F^{I}_{12} \, \gamma_3 \zeta  + \ex^{-\lambda} \slashed{\partial} \varphi^i \, \zeta + 2 \partial_{\varphi^i} \mc{W} \, \zeta  \, , \nn \\
\label{eq:app_VariationHyperino}
	0 &= \partial_\rho \mc{W} \, \zeta + \frac{1}{2} \ex^{-\lambda} \left( \partial_a \rho + 2\ii \partial_\rho Q_a \right) \gamma^a \zeta \, .
\end{align}
The first three are the same as \eqref{eq:app_SimplifiedEqs_2}, and the last one arises from the variation of the hyperino.

From these equations, using the same techniques that lead to \eqref{eq:app_SUSYBilinears_4d_1}--\eqref{eq:app_SUSYBilinears_4d_2} we find the following relations
\begin{align}
	\dd S &= 0 \, , \quad \dd( \ex^\lambda P) = - \frac{1}{4} \sum_I (X^I)^{-1} \xi \hook F^I \, , \quad \dd K = 0 \, , \nn\\
	\dd \xi^\flat &= 2 P \left( 1 + PS^{-1} \ex^\lambda \mc{W}  \right) \vol \, , \nn\\
	\dd ( \ex^\lambda S ) &= - \ex^{2\lambda} \frac{\mc{W}}{2} \, \xi \hook \vol \, ,\nn \\
	\dd X^I &= \sum_i \partial_{\varphi^i} X^I \, \dd \varphi^i = \frac{1}{2} \ex^{-\lambda} P^{-1}  \sum_i \varsigma_{Ii} X^I \sum_J \partial_{\varphi^i} (X^J)^{-1}  \, \xi \hook F^J \nn \\
	\label{eq:4d_dXI_Hyper}
	&= S^{-1} \ex^\lambda \sum_i \varsigma_{Ii} X^I \partial_{\varphi^i} \mc{W} \, \xi \hook \vol \, , 
	\end{align}
	and
	\begin{align}
	S \dd \rho &= 2 \ex^\lambda \partial_\rho \mc{W} \, \xi \hook \vol - P \sinh\rho * D\theta \, , \nn\\
	\label{eq:4d_xiDtheta_Hyper}
	\xi \hook D\theta &=  P \ex^\lambda \zeta_I X^I \, ,
\end{align}
as well as the algebraic condition
\begin{equation}
	\ex^{-\lambda} \sum_I (X^I)^{-1} F^I_{12} = -4 - PS^{-1} \ex^\lambda \mc{W} \, . 
\end{equation}
In these expression we have summarized \eqref{eq:4d_GaugeFixingSymplecticCondition} as 
\begin{equation}
	X^I = \exp \left( \sum_i \frac{1}{2} \varsigma_{Ii} \varphi^i \right) \, ,
\end{equation}
and used the definitions of $\mc{W}$ and $Q_\mu$ in \eqref{hhbbdef} and \eqref{superpottext}, respectively. Compared to the discussion in the previous appendix, we highlight that 
\eqref{eq:4d_xiDtheta_Hyper} are new equations
arising from \eqref{eq:app_VariationHyperino}.

Once again, we find that $F^I$ is the top form of an equivariantly closed polyform. Using \eqref{eq:4d_dXI_Hyper}, we have
\begin{align}
\label{eq:app_dX^IelambdaP}
	\dd ( X^I \ex^\lambda P) &= -\frac{1}{4} X^I \sum_J \left( 1 + \sum_i \varsigma_{Ii} \varsigma_{Ji} \right) (X^J)^{-1} \xi \hook F^J \nn \\
	&= - \xi \hook F^I \, ,
\end{align}
which implies that $\Phi^{F^I}$ in \eqref{eq:4d_STUHyper_PhiFI} is equivariantly closed.
 
We find that
\begin{align}
\label{eq:app_e3lambdaWS}
	\dd ( \ex^{3\lambda} \mc{W} S ) &= 3 \ex^{2\lambda} \mc{W} \, \dd (\ex^\lambda S) + \ex^{3\lambda} \left( \partial_\rho \mc{W} \, S \dd \rho + \sum_i \partial_{\varphi^i} \mc{W} S \, \dd\varphi^i \right) \nn \\
	&= \ex^{4\lambda} \mc{V}\, \xi \hook \vol + \frac{1}{2} P \ex^{3\lambda} \sinh^2\rho \sum_I \zeta_I X^I *D\theta \, ,
\end{align}
having used the definition \eqref{eq:4d_HyperSTU_Potential} of the scalar potential in terms of the real superpotential. In order to relate this to the integrand of the action in \eqref{eq:4d_HyperSTU_SoffShell}, we also need Maxwell's equations, which now include a matter term
\begin{equation}
	\dd \left[ (X^{I})^{-2} F^I_{12} \right] = - \zeta_I \ex^{2\lambda} \sinh^2 \rho * D\theta \, .
\end{equation}
Combining this with \eqref{eq:app_dX^IelambdaP} and \eqref{eq:app_e3lambdaWS}, we find
\begin{align}
	\dd \left( \ex^{3\lambda} \mc{W} S + \frac{1}{2}\ex^\lambda P \sum_I (X^I)^{-1} F^I_{12} \right) &= \xi \hook \left( \ex^{4\lambda} \mc{V} \, \vol - \frac{1}{2} \sum_I (X^I)^{-2} F^I_{12} \, F^I \right) \, ,
\end{align}
thus proving that $\Phi^S$ in \eqref{eq:4d_STUHyper_Phi} is equivariantly closed. 

Finally, we can show that there is a combination of the $\Phi^{F^I}$ that is equivariantly exact. This follows directly from \eqref{eq:4d_xiDtheta_Hyper} and the definition \eqref{dthetastext}:
\begin{equation}
	(\dd - \xi \hook\, ) D\theta = \zeta_I \Phi^{F^I} \, .
\end{equation}

\providecommand{\href}[2]{#2}\begingroup\raggedright\endgroup

\end{document}